\documentclass[lettersize, journal]{IEEEtran}

\usepackage{amsmath,amsfonts,dsfont}
\usepackage{amssymb,amsthm}
\usepackage{mathtools}
\usepackage{array}
\usepackage[caption=false,font=small,labelfont=sf,textfont=sf]{subfig}
\usepackage{textcomp}
\usepackage{stfloats}
\usepackage{url}
\usepackage{verbatim}
\usepackage{graphicx}
\usepackage{acronym}
\usepackage{cite}
\usepackage[bookmarks,colorlinks]{hyperref}

\usepackage[dvipsnames]{xcolor}

\usepackage{enumitem}

\usepackage[ruled,linesnumbered]{algorithm2e}
\SetKwInput{KwData}{\textbf{Init}} 

\definecolor{NewColor}{rgb}{0.2,0,0.5}

\newcommand{\myVec}[1]{{\boldsymbol{#1}}}
\newcommand{\myMat}[1]{{\boldsymbol{#1}}}
\newcommand{\mySet}[1]{\mathbb{#1}}
\newcommand{\E}{\mathds{E}}		 
\newcommand{\CovMat}[1]{\myMat{C}_{#1}} 
\newcommand{\opt}{^{\rm o}}	
\newcommand{\figwidth}{0.9\columnwidth}

\acrodef{adc}[ADC]{analog-to-digital converter}
\acrodef{aoa}[AoA]{angle-of-arrival}
\acrodef{af}[AF]{array factor}
\acrodef{cs}[CS]{compressed sensing}
\acrodef{dtft}[DTFT]{discrete-time Fourier transform} 
\acrodef{csi}[CSI]{channel state information}
\acrodef{bpsk}[BPSK]{binary phase shift keying}
\acrodef{map}[MAP]{maximum a-posteriori probability}
\acrodef{snr}[SNR]{signal-to-noise ratio}
\acrodef{bs}[BS]{base station}  
\acrodef{mimo}[MIMO]{Multiple-input multiple-output}
\acrodef{mse}[MSE]{mean-squared error}
\acrodef{pdf}[PDF]{probability density function}
\acrodef{rv}[RV]{random variable}
\acrodef{vm}[VM]{vector modulator}
\acrodef{evm}[EVM]{error vector magnitude}
\acrodef{hbf}[HBF]{hybrid analog/digital beamforming}
\acrodef{mmse}[MMSE]{minimum MSE}
\acrodef{fom}[FOM]{figure of merit}
\acrodef{enob}[ENOB]{effective number of bits}
\acrodef{vga}[VGA]{variable gain amplifier}
\acrodef{qaf}[QAF]{quadrature all-pass filter}

\theoremstyle{plain}
\newtheorem{theorem}{Theorem}

\newtheorem{lemma}[theorem]{Lemma}

\begin{document}

\title{Robust Task-Specific Beamforming with Low-Resolution ADCs for Power-Efficient Hybrid MIMO Receivers}
\author{
	\IEEEauthorblockN{Eyyup Tasci, Timur Zirtiloglu, Alperen Yasar, Yonina C. Eldar, Nir Shlezinger, and Rabia Tugce Yazicigil\\
	} 
	\thanks{ 
	    Parts of this work were presented in the 2022 IEEE International Conference on Acoustics, Speech, and Signal Processing (ICASSP) as the paper \cite{zirtiloglu2022power}. 
	    E. Tasci, T. Zirtiloglu, A. Yasar, and R. T.  Yazicigil are with the  Department of ECE, Boston University, Boston, MA (e-mail:\{etasci2, timurz, ayasar, rty\}@bu.edu). 
		N. Shlezinger is with the School of ECE, Ben-Gurion University of the Negev, Beer-Sheva, Israel (e-mail: nirshl@bgu.ac.il).
		Y. C. Eldar is with the Faculty of Math and CS, Weizmann Institute of Science, Rehovot, Israel (e-mail: yonina.eldar@weizmann.ac.il). 	\\
		E. Tasci and T. Zirtiloglu contributed equally to this work.
		}
		\vspace{-0.7cm}
}




\maketitle

\begin{abstract}
Multiple-input multiple-output (MIMO) systems exploit spatial diversity to facilitate multi-user communications with high spectral efficiency by beamforming. As MIMO systems utilize multiple antennas and radio frequency (RF) chains, they are typically costly to implement and consume high power. A common method to reduce the cost of MIMO receivers is utilizing less RF chains than antennas by employing hybrid analog/digital beamforming (HBF). However, the added analog circuitry involves active components whose consumed power may surpass that saved in RF chain reduction. An additional method to realize power-efficient MIMO systems is to use low-resolution analog-to-digital converters (ADCs), which typically compromises signal recovery accuracy. In this work, we propose a power-efficient hybrid MIMO receiver with low-quantization rate ADCs, by jointly optimizing the analog and digital processing in a hardware-oriented manner using task-specific quantization techniques. To mitigate power consumption on the analog front-end, we utilize efficient analog hardware architecture comprised of sparse low-resolution vector modulators, while accounting for their properties in design to maintain recovery accuracy and mitigate interferers in congested environments. To account for common mismatches induced by non-ideal hardware and inaccurate channel state information, we propose a robust mismatch aware design. Supported by numerical simulations and power analysis, our power-efficient MIMO receiver achieves comparable signal recovery performance to power-hungry fully-digital MIMO receivers using high-resolution ADCs. Furthermore, our receiver outperforms the task-agnostic HBF receivers with low-rate ADCs in recovery accuracy at lower power and successfully copes with hardware mismatches.
\end{abstract}


\section{Introduction}
\ac{mimo} technology is widely used in modern wireless communication systems due to their superior data capacity, improved coverage, and highly robust multi-user support \cite{key_tech_5G, navrati_5G_survey, shlezinger2018spectral}. 
Although the theoretical advantages of \ac{mimo} communications are well-established, practical trade-offs between power consumption, spectral efficiency, and reliability emerge in hardware implementations \cite{ naguib_from_theor, larsson_scaling, shlezinger2019asymptotic}. \ac{mimo} receivers utilize multiple signal acquisition chains for spatial signal processing, each consisting of a radio-frequency (RF) front end performing RF signal amplification with low noise, followed by down-conversion to baseband. These continuous-time analog baseband signals are translated into a digital representation for further processing. Analog-to-digital conversion is performed in two steps: the continuous-time analog signal is sampled into a discrete-time signal, and then quantized into a discrete-amplitude representation stored as digital bits \cite{eldar_2015}. This process is usually carried out in hardware using uniform scalar \acp{adc}~\cite{walden_1999}. \ac{mimo} systems traditionally apply fully-digital data acquisition with spatial signal processing using high-resolution quantization and Nyquist sampling rates, leading to high power consumption and hardware design complexity. 

A common approach to mitigate the increased cost of \ac{mimo} receivers is to utilize fewer RF chains and \acp{adc} than antennas via \ac{hbf}. Such architectures incorporate an additional analog combiner circuit before the acquisition, allowing dimensionality reduction~\cite{mendez2016hybrid, ioushua2019family} while preserving the ability of the \ac{mimo} array in achieving directed beamforming via, e.g., holographic techniques \cite{huang2020holographic}. In fact, \ac{hbf} is also utilized without RF chain reduction to boost pre-acquisition spatial interferer rejection \cite{soer_JSSC11_switchcap, soer_JSSC17_4beam, harish_procieee2016, golabighezelahmad20200}. Nonetheless, the introduction of an analog combiner comprised of active components may also be power-hungry. An additional power reduction technique is to use low-resolution acquisition, connecting each antenna to a low-quantization rate \ac{adc} \cite{heath_apr17, nossek_2018, alan_2018, liu_2017}. However, the distortion added by coarse quantization results in degraded signal recovery. 

Recently, a task-specific quantization framework was introduced to design \ac{mimo} receivers that combine both \ac{hbf} with bit-constrained \acp{adc} \cite{shlezinger2019hardware, neuhaus2020taskbased, shlezinger2019deep, salamatian_2019}. In task-specific quantization, the analog front end is designed to be aware of the use of low-resolution \acp{adc} and the desired task. Task-specific receivers combine the signals in analog such that the quantization distortion hardly affects the task information recovered in digital. Task-specific design yield improved signal recovery in \ac{mimo} communications \cite{shlezinger2019asymptotic, yonina_2021_meta, yonina_learning_2020} and radar~\cite{xi2020bilimo}. 

Despite its performance gains, implementing \ac{mimo} receivers utilizing task-specific quantization gives rise to challenges associated with the usage of a configurable analog combiner. First, such circuitry may rely on costly and power-hungry active components. Moreover, the task-specific design configures the analog front end based on the channel state information, e.g., knowledge of the \acp{aoa} of the desired signals. While this assumption is often adopted in adaptive beamforming systems \cite{soer_JSSC11_switchcap, soer_JSSC17_4beam, harish_procieee2016, golabighezelahmad20200, Zhang2018_RobustBeamforming}, it makes the system sensitive to \ac{aoa} inaccuracies. Current literature on tackling such mismatches via robust adaptive beamforming, e.g., \cite{Zhang2018_RobustBeamforming}, assumes usage of infinite-resolution \acp{adc}, and has not been employed in low-bit settings. Furthermore, in practice, analog circuitry, and particularly limited cost and configurable designs, induce mismatches, such that the gain and phase response of the analog combiner may differ from the ideal response. Such mismatches degrade task recovery performance of the system \cite{deepak2019impairments, bakr2009impairments}. The above challenges motivate a hardware-aware design of low-bit \ac{hbf} receivers, which accounts for low-power and non-ideal combiner circuitry while coping with \ac{aoa} mismatches. 

In this work, we study power-efficient hybrid \ac{mimo} receivers implementing bit-constrained signal recovery under \ac{aoa} mismatches and hardware non-idealities. We consider \ac{mimo} communications in congested environments, 
where the system task is to recover the desired signals and reject interferers. We present a hardware architecture utilizing low-resolution \acp{adc} and a programmable analog pre-processing front end. Since power consumption is highly implementation-dependent, we consider analog combiners implemented using \acp{vm}, which is a common circuitry for realizing configurable gain and phase~\cite{ellinger2010VM}. Since the analog power consumption is directly proportional to the hardware complexity \cite{cornell2020power}, we reduce the power by boosting the \acp{vm} to be either discrete or sparsely activated \cite{zirtiloglu2022power}. We propose a task-specific algorithm co-integrating these hardware-level techniques for accurate and power-efficient signal recovery. 

Next, we propose a robust counterpart of the task-specific design that co-optimizes power-efficient recovery in the presence of mismatches. We tackle the presence of \ac{aoa} errors by proposing a robust \ac{hbf} design and cope with hardware non-idealities by developing a hardware-compliant model, which is incorporated into the design procedure. We compare the proposed system with task-agnostic \ac{mimo} systems in terms of signal recovery accuracy, receiver beam pattern, and power consumption. For the latter, we use power profiles derived from the measured power consumption of state-of-the-art hardware implementations ~\cite{soer_JSSC17_4beam, ellinger2009vm, ragaie09mixer, rebeiz17mixer, mendez2016hybrid, lee10badc, sodini2008ADC}. Our comparative study shows that at a significantly reduced quantization rate, our design achieves accurate signal recovery comparable to the performance of fully-digital \ac{mimo} receivers using high-resolution \acp{adc} while being robust to practical mismatches. Furthermore, the task-specific receiver notably outperforms task-agnostic architectures operating under similar bit constraints. Regarding beam pattern, we demonstrate that our design attenuates the interferers by $\ge30$dB while still being able to provide sufficient gain for desired signals. The task-specific receiver reduces the power consumption by at least $58\%$ compared to task-agnostic fully-digital \ac{mimo} and \ac{hbf} receivers. 


\begin{figure}
    \centering
    \includegraphics[width=\columnwidth]{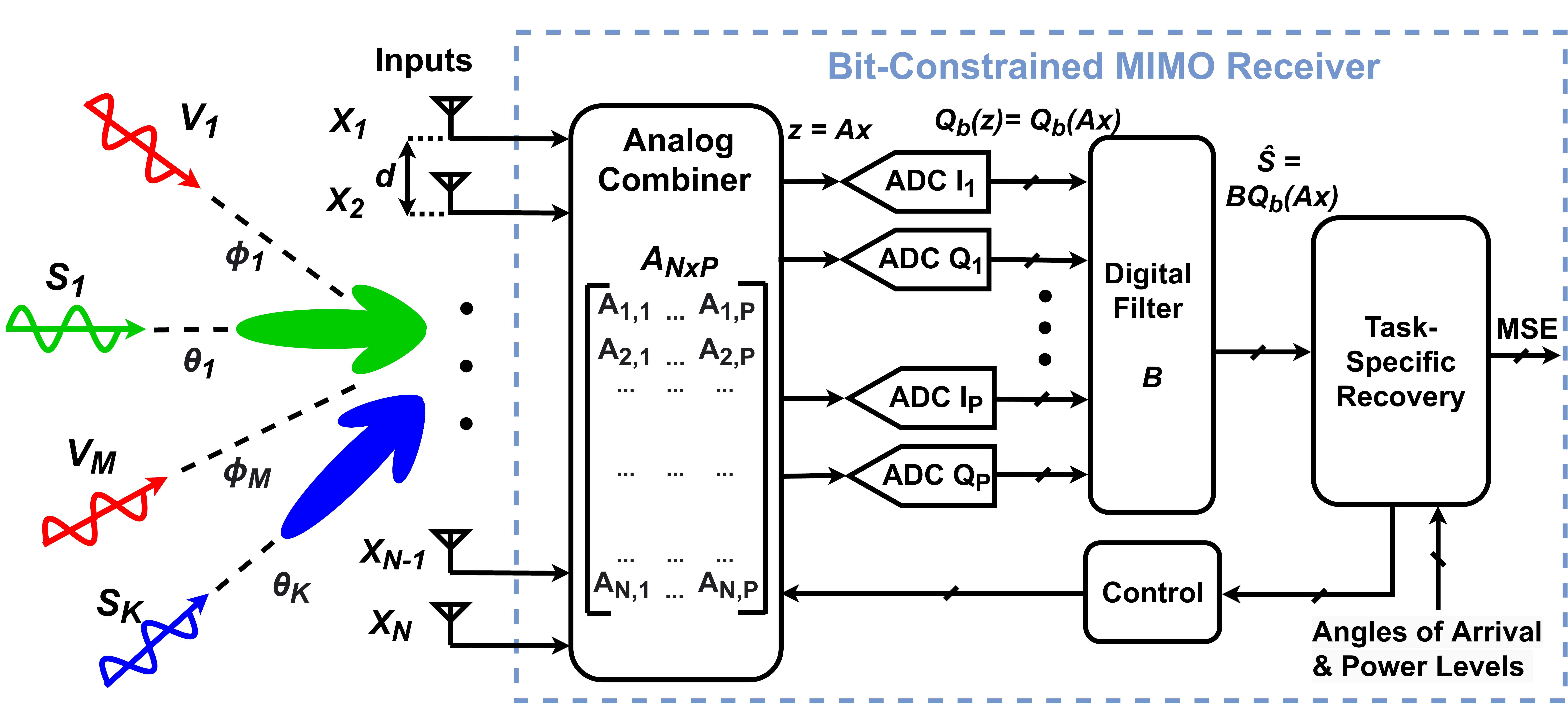}
    \caption{Task-specific hybrid MIMO receiver system.}
    \label{fig:System1}
\end{figure}

The rest of the paper is organized as follows: Section~\ref{sec:Model} reviews the system model and presents the receiver architecture. The non-mismatched \ac{hbf} design algorithm is provided in Section~\ref{sec:Non-MismatchedTS}, and its robust counterpart is derived in Section~\ref{sec:Mismatches}. Section~\ref{sec:Analysis} analyzes our model-based performance evaluation supported by numerical simulations and power consumption estimates, and Section~\ref{sec:Conclusions} concludes the paper.

Throughout the paper, we use boldface lower-case letters for vectors and boldface upper-case letters for matrices. The $n$th element of a vector $\myVec{x}$ and the $(n,m)$th element of a matrix $\myMat{X}$ are denoted as $[\myVec{x}]_n$ and $[\myMat{X}]_{n,m}$, respectively. We use $j$, $(\cdot)^H$, $\E\{\cdot \}$, $\rm Re\{\cdot\}$, and $\rm Im\{\cdot\}$ for the imaginary unit, Hermitian transpose, stochastic expectation, real and imaginary parts. Finally, $\myMat{I}_n$ is the $n\times n$ identity matrix, and $\mySet{C}$ is the set of complex numbers.
\section{System Model}
\label{sec:Model}
In the following, we first review the signal model relating the observed signals and the task vector, hardware implementation of our design, and possible system mismatches in Subsection~\ref{ssec:ModelHybrid}. Then, based on these models, we formulate the system design problem in Subsection~\ref{ssec:ModelProblem}.

\subsection{Hybrid MIMO Receiver with Embedded Beamforming}
\label{ssec:ModelHybrid}
We study low-power hybrid \ac{mimo} receivers, focusing on the task of recovering desired signals under the presence of interferers via beamforming. Our description of the system operation commences with the received signals, after which we model the operations of the \acp{adc}, analog pre-processing, and common non-idealities that are likely to be encountered in such settings. 

\subsubsection{Signal Model} Consider a hybrid \ac{mimo} receiver with $N$ antenna elements and $P$ RF chains. Two \acp{adc} are needed in each RF chain to quantize down converted in- (I) and quadrature-phase (Q) baseband signals, resulting in $2 \cdot P$ \acp{adc} in total. {All signals are assumed to be sampled above Nyquist sampling rate, yet the \acp{adc} have finite rate quantizers.} Let $\myVec{x} = [x_{1}, \ldots, x_{N}]$ be the vector signal observed at the $N$ antenna elements. The received signal $\myVec{x}$ incorporates a set of $K$ desired signals denoted $[s_1,..., s_K]$, received from sources at relative angles $[\theta_1, \ldots, \theta_K]$, respectively. The received signal also includes $M$ interferers denoted $[v_1, \ldots, v_M]$, from sources at relative angles $[\phi_1, \ldots, \phi_M]$, respectively. We assume that the receiver has the prior knowledge of \acp{aoa} of desired signals and interferers. All sources are assumed to be narrowband Gaussian signals lying at the far-field, and thus the  received signal is 
\begin{equation}
\label{eq:observation}
    \myVec{x} = \sum_{k=1}^{K} s_k\myVec{a}(\theta_k) + \sum_{m=1}^{M} v_m\myVec{a}(\phi_m) + \myVec{w}.
\end{equation}
In  \eqref{eq:observation}, $\myVec{w}$ is additive white Gaussian noise with variance $\sigma_w^2$, and $\myVec{a}(\psi) \in \mySet{C}^{N}$ is the steering vector whose entries are 
\begin{equation}
    [\myVec{a}(\psi)]_n = e^{-j2\pi n \frac{d}{\lambda}\sin{\psi}},
\end{equation}
where $d$ is the spacing between antenna elements and $\lambda$ is the wavelength of the received signals. An illustration of this system is depicted in Fig. \ref{fig:System1}. 

Define the steering matrices $\myMat{M}_{\theta} \in \mySet{C}^{N \times K}$ and $\myMat{M}_{\phi} \in \mySet{C}^{N \times M}$ such that $[\myMat{M}_\theta]_{n,k} = [\myVec{a}(\theta_k)]_n.$ and $[\myMat{M}_\phi]_{n,m} = [\myVec{a}(\phi_m)]_n$, and let $\myVec{s} = [s_1, ..., s_N]$ and $\myVec{v} = [v_1, ..., s_M]$ be the desired signal and interferer vectors, respectively. We can then write \eqref{eq:observation} as
\begin{equation}
\label{eq:observationMatrix}
    \myVec{x} = \myMat{M}_{\theta}\myVec{s} + \myMat{M}_{\phi}\myVec{v} +  \myVec{w}.
\end{equation}
By letting $\CovMat{\myVec{s}}$ and $\CovMat{\myVec{v}}$ be the covariance matrices of $\myVec{s}$ and $\myVec{v}$, respectively, it follows from \eqref{eq:observationMatrix} that the covariance matrix of the received signal $\myVec{x}$ is
\begin{subequations}
\label{eqn:Autocorr}
\begin{equation}
    \label{eqn:Autocorr1}
    \CovMat{\myVec{x}} = \myMat{M}_{\theta}\CovMat{\myVec{s}}\myMat{M}_{\theta}^H + \myMat{M}_{\phi}\myMat{C}_{\myVec{v}}\myMat{M}_{\phi}^H + \sigma_w^2\myMat{I}_N,
\end{equation}
and its correlation with the task of interest $\myVec{s}$ is given by 
\begin{equation}
    \label{eqn:Autocorr2}
    \CovMat{\myVec{s}\myVec{x}} = \CovMat{\myVec{s}}\myMat{M}_{\theta}^H. 
\end{equation}
\end{subequations} 

\subsubsection{Signal Acquisition}
We focus on hybrid \ac{mimo} receivers operating with low-resolution \acp{adc}. Here, the acquisition of  $\myVec{x}$ includes pre-processing in the analog domain followed by its down-conversion and quantization into digital form.

Let $\myMat{A} \in \mySet{C}^{P \times N}$ be the analog combiner matrix that performs analog pre-processing. The observed signal is processed in analog first, generating the $P \times 1$ representation of the analog combiner output $ \myVec{z} = [z_{1}, ..., z_{P}]$, which is forwarded to the \acp{adc}. Hence, the output vector $\myVec{z}$ is given by
\begin{equation}
\label{eq:z_eq_Ax}
    \myVec{z} = \myMat{A}\myVec{x}.
\end{equation}
The acquisition thus includes the conversion of $\myVec{z}$ into digital by the \acp{adc}, and the setting of the analog combiner $\myMat{A}$.

\begin{figure}
    \centering
    \includegraphics[width=\columnwidth]{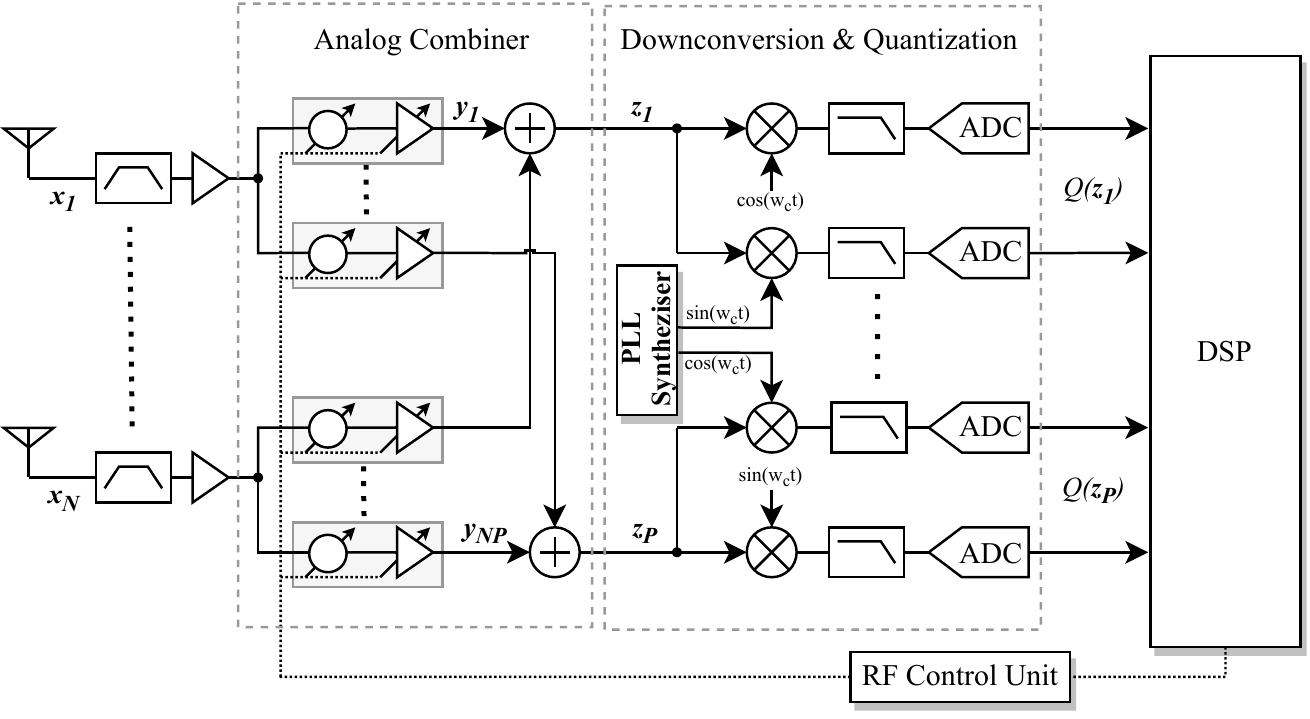}
    \caption{Schematic of a hybrid MIMO hardware implementation.}
    \label{fig:HardwareSystem}
\end{figure}

\textbf{Analog Hardware:} A high level hardware architecture implementing a hybrid \ac{mimo} system is depicted in Fig.~\ref{fig:HardwareSystem}. The analog combiner $\myMat{A}$ represents the adjustable analog pre-processing carried out on the received signals prior to their conversion to digital. The complex multiplication that each element of $\myMat{A}$ performs can be realized by means of serial phase shifters and \acp{vga}. Another alternative that is commonly used for implementing reconfigurable analog processing, which is considered in this work, utilizes \acp{vm} with low noise. Here, each complex entry of $\myMat{A}$ corresponds to a single \ac{vm} component, thus, for the $N \times P$ analog combiner matrix $\myMat{A}$, a total of $N\times P$ \acp{vm} are utilized. Signals are summed and downconverted at the output producing $2 \cdot P$ baseband output voltages, which are used as inputs to the \acp{adc}. 

\textbf{VMs:} A \ac{vm} is an analog circuitry that applies different variable gain and polarization to each of the I/Q components of a signal to realize gain and phase shift. The mapping carried out by each \ac{vm} is constrained to take values in a discrete set $\mathcal{A}$. To accommodate the ability to deactivate a \ac{vm}, we assume that $0 \in \mathcal{A}$. The resulting model for a \ac{vm}-based analog combiner, where the entries of $\myMat{A}$ take values in $\mathcal{A}$, accommodates various architectures, such as phase-shifter-based combiners \cite{mendez2016hybrid}, for which $\mathcal{A}$ is the union of the unit circle and the origin. 

\begin{figure}
\centering
  \subfloat[]{%
       \includegraphics[width=\linewidth]{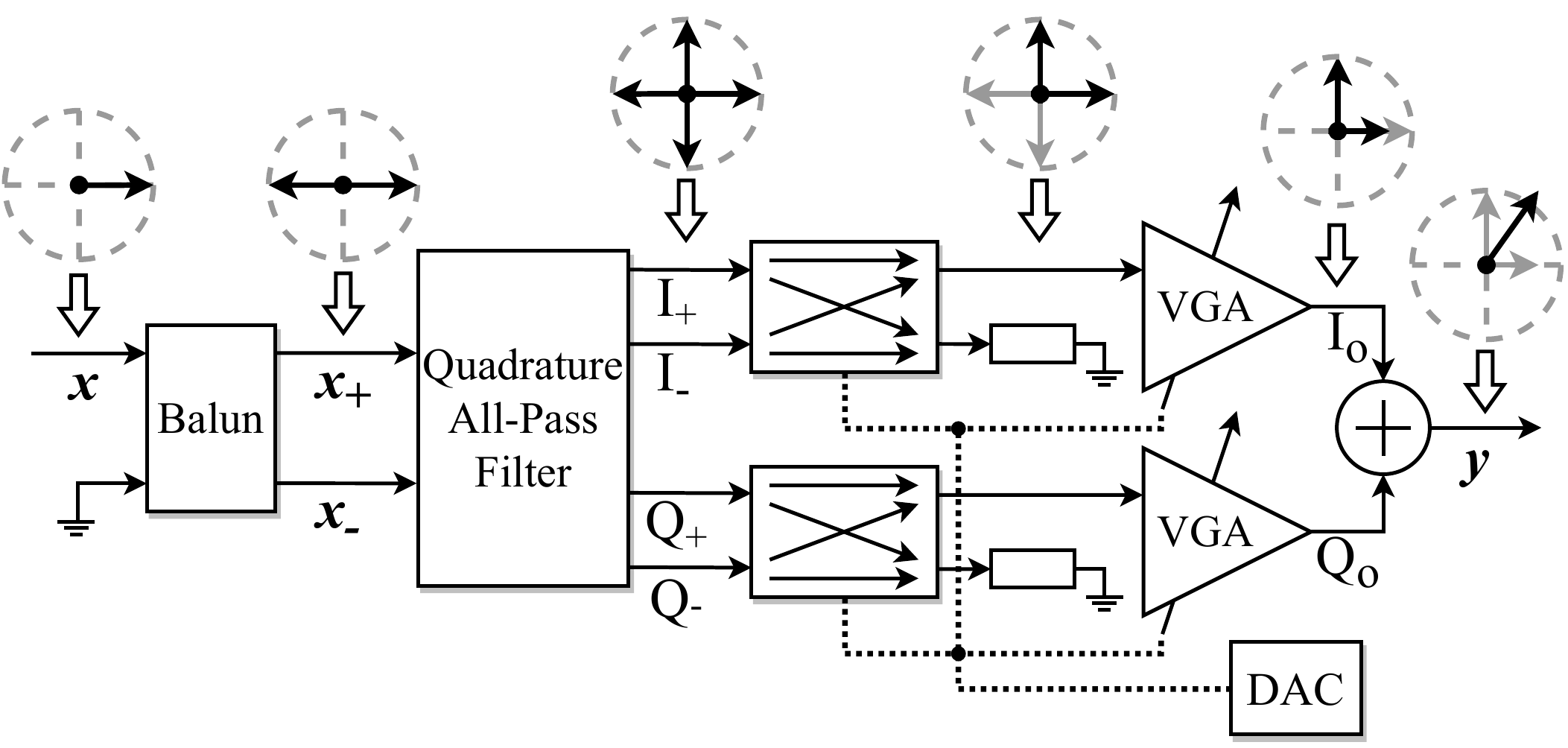}
       \label{fig:VM_architecture}}
    \hfill
  \subfloat[]{%
        \includegraphics[width=0.5\linewidth]{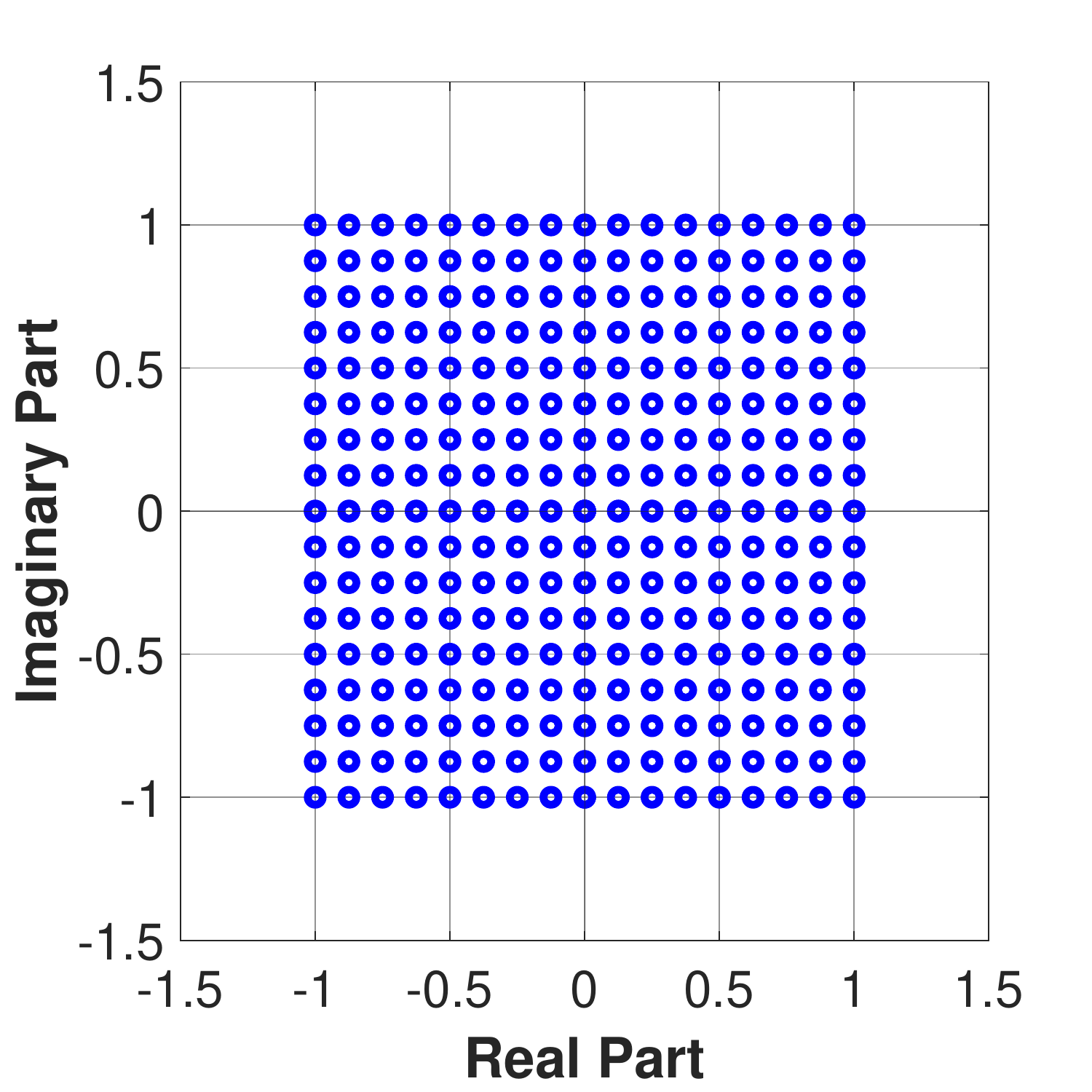}
        \label{fig:VM_constDiagram}}
  \caption{(a) \ac{vm} architecture and (b) constellation diagram of a 4-bit \ac{vm}.} 
  \label{fig:VM_2} 
\end{figure}

Since such analog processing can be costly in terms of power, one often utilizes Cartesian \acp{vm}, known to be highly power efficient \cite{Joram2009_VM}. 
The building blocks of a Cartesian \ac{vm} are illustrated in Fig.~\ref{fig:VM_architecture}. Here, the \ac{vm} input is first converted to its differential parts by a balun. The differential input is then fed into a \acl{qaf}, yielding four orthogonal components with a phase shift of $0^\circ$, $180^\circ$, $90^\circ$, and $270^\circ$ in its output. The quadrant of the final vector is determined by selecting the corresponding I/Q paths, while other unused paths are terminated with matched impedance. The desired complex gain can be realized by properly weighting the selected orthogonal components in magnitude using \acp{vga}. The control signals of the \acp{vga} are generated by the $r$-bit digital output of the digital-to-analog converter, which determines a discrete set $\mathcal{A}$ and the resolution of the Cartesian \ac{vm}. Higher resolution of complex gain/phase states comes at the cost of higher power consumption. The vectors generated at the output of each block are illustrated at the top of Fig.~\ref{fig:VM_architecture}. For the $r$-bit Cartesian \ac{vm} architecture, the set $\mathcal{A}$ is given as
\begin{equation}
    \mathcal{A} = \left\{g + j h \mid g, h \in \left\{ 0, \pm \frac{2}{2^r}, \pm \frac{4}{2^r}, \dots, \pm 1 \right\}\right\}.
    \label{eqn:CartA}
\end{equation}
Fig.~\ref{fig:VM_constDiagram} illustrates the normalized complex constellation of all available complex gain/phase states for a 4-bit \ac{vm}. 

\textbf{ADCs:} The analog combiner output $\myVec{z}$ is converted into a digital representation using $2 \cdot P$ identical uniform \acp{adc}, each with $b$ levels. The overall number of bits used for representing $\myVec{x}$ in digital is thus $ 2\cdot P\lceil \log_2b \rceil$. The resulting vector processed in the digital domain is $\mathcal{Q}_b (\myVec{z})$, where $\mathcal{Q}_b(\cdot)$ is the element-wise uniform quantization operator with $b$ levels. 

\subsubsection{Common Non-Idealities} 
Hybrid \ac{mimo} beamforming is often carried out under hardware and environmental non-idealities. In the following, we evaluate our system robustness under two types of mismatches: $(1)$ inaccurate knowledge of the \acp{aoa} of the desired users; and $(2)$ hardware non-idealities stemming from phase and gain errors in the \acp{vm}. 

\textbf{Inaccurate \acp{aoa}:}
A common mismatch encountered in \ac{mimo} beamforming occurs when the receiver has a noisy estimate of the true \acp{aoa} of the desired signals $\myVec{\theta} = [\theta_1, \ldots, \theta_K]$. This may arise due to the limited resolution of the \ac{aoa} estimation sensors and localization errors. Moreover, a malicious adversary can attack a task-specific sensor by corrupting the estimated \acp{aoa} provided to the system.
To incorporate such mismatches into our system model, we represent the \acp{aoa}  $\myVec{\theta}$ by a noisy estimate with an error margin. Assuming the mismatches of the \acp{aoa} are limited to some degree $\epsilon > 0$, we define an interval $\Theta(\myVec{\theta}, \epsilon)$ that contains the true \acp{aoa} as: 
\begin{equation}
    \Theta(\myVec{\theta}, \epsilon) = \{\myVec{\bar{\theta}} \mid \|\myVec{\bar{\theta}} - \myVec{\theta}\|_{\infty} \leq \epsilon\}.
\end{equation}

\textbf{Phase/Gain Mismatches in the Analog Hardware:} 
The analog pre-processing hardware has 
impairments in practice. In this work, we consider linear hardware impairments, including the phase and gain mismatches between the I/Q paths in \acp{vm}. Mismatches in the circuitry of quadrature all-pass filters or in \acp{vga} in \acp{vm} typically results in imbalances in the amplitude and phase of the output of the analog pre-processing, i.e., in $\myVec{z}$. Such mismatches can occur within a single \ac{vm}, as well as between different \acp{vm}, severely impacting the recovery accuracy if they are not calibrated. Under phase and gain mismatches, the matrix $\myMat{A}$ cannot perform accurate complex multiplication, hence the set $\mathcal{A}$ becomes corrupted. We denote this corrupted set by $\bar{\mathcal{A}}$. Hardware-compliant mathematical modeling of $\bar{\mathcal{A}}$ is proposed in Section~\ref{sec:Mismatches}.

\subsection{Problem Formulation}
\label{ssec:ModelProblem}
Our goal is to tune the analog combiner $\myMat{A}$ based on the above signal and hardware model. Our design is optimized for multiple tasks simultaneously, including accurate signal recovery, spatial interferer suppression, and power efficiency. We aim for the task-specific \ac{mimo} receiver to perform robust recovery when deployed under noisy estimation of the \acp{aoa} and hardware non-idealities. 

\textbf{Signal Recovery:} The main task of the receiver is to recover the
desired signal $\myVec{s}$ from the digital representation $\mathcal{Q}_b (\myVec{z})$. Our design objective is the \ac{mse}, being widely used for evaluating signal recovery, and is closely related to the \ac{evm} measure utilized in evaluating acquisition hardware. The \ac{mse} is defined by
\begin{equation}
\label{eqn:ObjectiveA}
    \textrm{MSE} := \E\{\|\myVec{s}- \E\{\myVec{s}|\mathcal{Q}_b (\myVec{z})\}\|^2\}.    
\end{equation}

\textbf{Interferer Suppression:} The MSE objective focuses only on the
ability to recover $\myVec{s}$. As such, it may prefer settings of $\myMat{A}$ in which the
effect of $\myVec{v}$ is mitigated in digital. In practice, it is often preferable to reject interferers in the analog domain, since strong spatial interferers may lead to receiver desensitization and increased dynamic range requirement of the \acp{adc} \cite{Koh2007_dynamicRange}. Consequently, we explicitly require the analog combiner to suppress the spatial interferers. Since the contribution of $\myVec{v}$ on the analog combiner output $\myVec{z}$ takes the form $\myMat{A}\myMat{M}_{\phi}$, we penalize interferer rejection via the max norm of $\myMat{A}\myMat{M}_{\phi}$, i.e.,
\begin{equation}
\label{eqn:ObjectiveMax}
   {\rm IntRej}(\myMat{A}) :=  \|\myMat{A}\myMat{M}_{\phi}\|_{\rm max} = \max_{i,j}\big( |[\myMat{A}\myMat{M}_{\phi}]_{i,j}|\big).
   \vspace{-0.1cm}
\end{equation} 

\textbf{Power Efficiency:} The power consumption of the MIMO receiver front end is dictated by the individual cost of each hardware component, including local oscillator generators (LO Gen), low-noise amplifiers, mixers, filters, and \acp{adc}. Under the commonly employed \ac{adc} performance measure Walden \ac{fom} \cite{walden_1999}, $P_{\rm ADC}/(f_s \cdot 2^{\rm ENOB})$, the performance of the \ac{adc} becomes proportional to the power consumption $P_{\rm ADC}$, inversely proportional to the sampling rate, $f_s$, and the \ac{enob}, which is approximately proportional to the number of levels $b$ \cite{YazicigilJSSC-ADCPower}.

The power consumption of an analog combiner hardware $\myMat{A}$ is highly dependent on two factors: $1)$ how many different values can its entries take, i.e., $|\mathcal{A}|$. This quantity is dictated by the resolution of \acp{vm}; and $2)$ which of its entries are active, namely, the sparsity level of $\myMat{A}$. If the sparsity level of $\myMat{A}$ increases while reducing its resolution, i.e., using a coarse $\myMat{A}$, the receiver power consumption is significantly reduced.

\textbf{Robustness to Mismatches:}
As we show in Section~\ref{sec:Non-MismatchedTS}, the \ac{mse} objective can be formulated as a function of the \acp{aoa} of the desired signals, which are assumed to be known up to some degree $\epsilon$. However, tuning the system using an inaccurate estimation of the \acp{aoa} may lead to high recovery errors. To address this problem, the analog combiner matrix $\myMat{A}$ and the digital processing should be designed by incorporating the range of possible \acp{aoa}, $\Theta(\myVec{\theta}, \epsilon)$, into the \ac{mse} objective, where the maximal tolerable error $\epsilon$ is assumed to be known.

Under gain and phase imbalances due to hardware impairments, the analog combiner $\myMat{A}$ does not perform the ideal complex multiplication that it is tuned for. Instead, it performs a complex multiplication corresponding to one of the values in the corrupted discrete set $\mathcal{\bar A}$. We aim to make our system robust against these hardware mismatches by requiring that the elements of the analog combiner matrix $\myMat{A}$ can only be configured to the values of the set $\mathcal{\bar A}$.
%
\section{Non-Mismatched Low-Power HBF}
\label{sec:Non-MismatchedTS}
The \ac{hbf} design problem detailed in Subsection~\ref{ssec:ModelProblem} accounts for several practical considerations. We initially consider that the system operates without the hardware mismatches and inaccurate AoAs to tackle this challenging design problem gradually. Namely, our goal here is to propose a design scheme which jointly accounts for the requirements of signal recovery, power consumption, and analog interference rejection. We consider the case where the number of ADCs ($P$), their resolution ($b$), and the values that the \acp{vm} can be configured to ($\mathcal{A}$) are dictated by the hardware performance requirements, and optimize $\myMat{A}$ accordingly. In particular, we first focus on signal recovery in Subsection~\ref{ssec:tbquant}, then we incorporate interferer suppression and power reduction in Subsection~\ref{ssec:intRej}. The resulting design is summarized in Subsection~\ref{subsec:Design}, and its properties are discussed in Subsection~\ref{subsec:Discussion}.

\subsection{Signal Recovery via Task-Specific Quantization}
\label{ssec:tbquant}
We formulate the recovery error of the desired signal $\myVec{s}$ from its estimate  $\hat{\myVec{s}} = \E\{\myVec{s}|\mathcal{Q}_b (\myVec{z})\}$. For any quantized representation of $\hat{\myVec{s}}$, the orthogonality principle implies that \eqref{eqn:ObjectiveA} can be decomposed as
\begin{equation}
\label{eqn:MSEdecomposed}
    {\rm MSE} = \E\{\|\myVec{s}-\hat{\myVec{s}}\|^2\} = \E\{\|\myVec{s}-\tilde{\myVec{s}}\|^2\} + \E\{\|\tilde{\myVec{s}}-\hat{\myVec{s}}\|^2\}, 
\end{equation}
where $\tilde{\myVec{s}}$ is \ac{mmse} estimate of the desired signal $\myVec{s}$ (which is a linear estimate for the model in \eqref{eq:observation} since the sources are Gaussian). The recovery error in \eqref{eqn:MSEdecomposed} is given by the sum of the \ac{mmse} and the distortion with respect to the \ac{mmse} estimate $\tilde{\myVec{s}}$. The second term depends on the configuration of the analog pre-processing, whereas the first term does not. One can thus use the excess \ac{mse} with respect to $\tilde{\myVec{s}}$, which depends on $\myMat{A}$ and is denoted henceforth as ${\rm ExMSE}(\myMat{A})$, as the signal recovery measure.

Configuring $\myMat{A}$ which minimizes \eqref{eqn:MSEdecomposed} is a special case of the task-specific (task-based) quantization setup studied in \cite{shlezinger2019hardware}. There, it was shown that while minimizing \eqref{eqn:MSEdecomposed} is likely to be analytically intractable, one can obtain accurate signal recovery by modeling the \acp{adc} as implementing non-subtractive dithered quantization while aiming to recover the \ac{mmse} estimate of $\myVec{s}$. By defining $\myMat{\Gamma} \triangleq \mathbf{C_{\myVec{s}\myVec{x}}} \mathbf{C_{\myVec{x}}}^{-1}$, the \ac{mmse} estimate can be written as $\tilde{\myVec{s}} = \myMat{\Gamma} x$. Since the MMSE estimate is linear, we let the digital processing outputs an estimate of the form $\hat{\myVec{s}} = \myMat{B} \mathcal{Q}_b(\myMat{A}\myVec{x})$ for some $\myMat{B}\in\mathbb{C}^{K\times P}$. In \cite{shlezinger2019hardware}, it is discussed that the choice of linear MMSE estimate does not affect the overall performance significantly because $\tilde{\myVec{s}}$ is the linear MMSE estimate of the desired signal. This holds true especially when the quantization error is small.

To formulate the \ac{mse} objective under the above considerations, set $\kappa := \eta^2(1 -\frac{\eta^2}{3b^2})^{-1}$ where $\eta$ is a coefficient tuned to guarantee negligible overloading probability of the \acp{adc} by determining the ratio of the \ac{adc} dynamic range to the standard deviation of its input. For instance, by setting $\eta=3$ one guarantees overloading with probability of at most $11\%$ for arbitrary signals by Chebyshev's inequality, and of less than $1\%$ for Gaussian signals. We can now reformulate the \ac{mse} in \eqref{eqn:MSEdecomposed} as stated in the following lemma (adapting \cite[Lem. 1]{shlezinger2019hardware} to complex signals):
\begin{lemma}
\label{lem:mse}
 When the \acp{adc} utilize non-subtractive dithered quantizers with vanishing overloading probability, the excess \ac{mse} in the objective \eqref{eqn:MSEdecomposed} becomes
 \begin{align}
      {\rm ExMSE}(\myMat{A}) \!= \! {\rm Tr}& \Bigg( \myMat{\Gamma} \CovMat{\myVec{x}}\myMat{\Gamma}^H\! -\! \myMat{\Gamma}\CovMat{\myVec{x}} \myMat{A}^H\!\bigg(  \myMat{A}\CovMat{\myVec{x}} \myMat{A}^H\notag \\
      &+ \!\frac{{2 \kappa \cdot {\rm Tr}(\myMat{A}\CovMat{\myVec{x}} \myMat{A}^H)}}{{3b^2 \cdot P}}{{\myMat{I}}_P} \bigg)^{ - 1}\! \myMat{A}\CovMat{\myVec{x}}\myMat{\Gamma}^H \Bigg).
      \label{eqn:ObjectiveA2}
 \end{align}
 This \ac{mse} is achieved by setting the digital filter to be 
		\begin{equation}
		\label{eqn:OptimalDesB}
		\myMat{B}\left( \myMat{A}\right) \!=\! {\myMat{\Gamma}} \CovMat{\myVec{x}} \myMat{A}^H \bigg( \myMat{A}\CovMat{\myVec{x}} \myMat{A}^H \!+ \!\frac{{2 \kappa \cdot {\rm Tr}(\myMat{A}\CovMat{\myVec{x}} \myMat{A}^H)}}{{3b^2 \cdot P}}{{\myMat{I}}_P} \bigg)^{ - 1}.
		\end{equation}
\end{lemma}
 
The digital matrix $\myMat{B}$ in Lemma~\ref{lem:mse} is the linear \ac{mmse} estimation matrix given $\mathcal{Q}_b(\myMat{A}\myVec{x})$. While Lemma~\ref{lem:mse} rigorously holds under some limiting assumptions on the system operation, i.e., using non-subtractive dithered quantizers, it also approximately holds when these assumptions are not satisfied for a broad range of input signals \cite{shlezinger2019hardware}.
In the following, we exploit this representation of the \ac{mse} objective to incorporate additional design considerations such that $\myMat{A}$ is optimized while meeting the performance requirements detailed in Subsection~\ref{ssec:ModelProblem}.

\subsection{Low-Power Interferer Suppression}
\label{ssec:intRej}
The objective \eqref{eqn:MSEdecomposed} admits a closed-form minimizer, see \cite[Thm. 1]{shlezinger2019hardware}. However, such a design only considers the signal recovery task and does not impose any structure on $\myMat{A}$. To account for the interference rejection requirement and to balance the power consumption of the analog circuitry, we formulate our design objective as
\begin{equation}
\label{eqn:objectiveFun}
    \mathcal{L}(\myMat{A}) = \textrm{ExMSE}(\myMat{A}) + \gamma_I\textrm{IntRej}(\myMat{A}) + \gamma_S \|\myMat{A}\|_{1,1}.
\end{equation}
In \eqref{eqn:objectiveFun}, $\|\cdot\|_{1,1}$ is the entry-wise $\ell_1$ norm operator, encouraging the design to deactivate multiple \acp{vm}. The usage of the $\ell_1$ norm can be interpreted as a convex relaxation of explicitly imposing sparsity via $\ell_0$-minimization \cite[Ch. 2]{foucart2013invitation}. 
The term $\textrm{IntRej}(\cdot)$ defined in \eqref{eqn:ObjectiveMax} boosts the suppression of spatial interferes, and it is convex since it is defined via the maximum norm.
Finally, the hyperparameters $\gamma_I, \gamma_S> 0$ are regularization coefficients, balancing the contribution of signal recovery \ac{mse}, spatial interferer rejection, and sparsity level of the analog combiner in the overall loss measure $\mathcal{L}(\myMat{A})$. 


The final consideration which is not accounted for in \eqref{eqn:objectiveFun} is the usage of discretized \acp{vm}. The resulting optimization is 
\begin{equation}
\label{eqn:OptProblem}
    \myMat{A}\opt = \underset{\myMat{A}\in \mathcal{A}^{P \times N}}{\arg \min} \mathcal{L}(\myMat{A}).
\end{equation}
The fact that the optimization problem is formulated over a discrete (i.e., non-convex) search space makes obtaining $\myMat{A}$ that optimizes the objective within $\mathcal{A}$ extremely challenging.
Nonetheless, one can utilize discrete optimization to recover suitable designs of $\myMat{A}$, as proposed next.

\subsection{Analog Combiner Design}
\label{subsec:Design}
Since the optimization problem \eqref{eqn:OptProblem} seeks to minimize its objective over a discrete set, we tackle it using projected gradient based optimization. Our  design strategy is comprised of $k_{\max}$ iterations for minimizing $\mathcal{L}(\myMat{A})$ over $\mySet{C}^{P\times N}$, with periodic projections onto the discrete $\mathcal{A}$. We use $\mathcal{O}_\mathcal{L}(\cdot)$ to denote the iterative optimizer with loss measure $\mathcal{L}$. 

By dividing the loss \eqref{eqn:objectiveFun} into a {\em task} term $\tilde{\mathcal{L}}(\myMat{A}) := {\rm ExMSE}(\myMat{A}) +  \gamma_I {\rm IntRej}(\myMat{A})$ (measuring signal recovery and interference rejection performance) and a {\em prior} term $\gamma_S \|\myMat{A}\|_{1,1}$ (boosting sparsity of the analog combiner), a candidate iterative method is proximal gradient descent. Here, each iteration first computes a gradient step in $\tilde{\mathcal{L}}(\myMat{A})$ with step-size $\mu>0$, and then takes the proximal mapping with respect to $\gamma_S \|\myMat{A}\|_{1,1}$. The resulting iterative procedure is \cite[Ch. 3]{foucart2013invitation}:

\begin{align}
&\mathcal{O}_{\mathcal{L}}(\myMat{A}) 
\!=\! \mathop{\arg\min}\limits_{\tilde{\myMat{A}}\in\mySet{C}^{P\times N}} \gamma_S \|\tilde{\myMat{A}}\|_{1,1} \!+\!\frac{1}{2} \left\|\tilde{\myMat{A}} \!-\! \big(\myMat{A} \!-\! \mu \nabla_{\myMat{A}} \tilde{\mathcal{L}}(\myMat{A}) \big)\right\|_{2,2}^2 \notag \\
&\quad=\mathcal{T}_{\gamma_S }\left\{\myMat{A} \!-\! \mu \nabla_{\myMat{A}} \left({\rm ExMSE}(\myMat{A}) \!+\!  \gamma_I {\rm IntRej}(\myMat{A})\right)\right\}.   
\label{eqn:OptConvex}  
\end{align}
Here, $\mathcal{T}$ is the element-wise complex soft-thresholding operator, given by $\mathcal{T}_{\lambda}(z):=e^{j \arg(z)}\max(|z|-\lambda,0)$. Note that both the \ac{mse} in \eqref{eqn:ObjectiveA2} and the interference rejection penalty in \eqref{eqn:ObjectiveMax} are differentiable in $\myMat{A}$, and the gradient can be computed using automatic differentiation engines, e.g., Autograd~\cite{maclaurin2015autograd}.

Every $k_{\rm proj}$ iterations, the intermediate $\myMat{A}^{(k)}$ is projected to account for the discrete \acp{vm} via the element-wise projection operator $\mathcal{P}_{\mathcal{A}}(z):=\arg\min_{a\in \mathcal{A}}\|a-z\|_2$. In particular, for the $r$-bit Cartesian \ac{vm} architecture, the projection operator boils down to element-wise uniform quantization, namely,
\begin{equation}
\label{eqn:Proj}
    \big[\mathcal{P}_{\mathcal{A}}(\myMat{A}) \big]_{p,n} \!=\! \mathcal{Q}_{2^r + 1}\!\left(\big[{\rm Re}\{\myMat{A}\} \big]_{p,n}\! \right) \!+\! j \mathcal{Q}_{2^r + 1}\!\left(\big[{\rm Im}\{\myMat{A}\} \big]_{p,n}\! \right),
\end{equation}
with $\mathcal{Q}_{b}(\cdot)$ being the $b$ level uniform mid-tread quantizer (see Subsection~\ref{ssec:ModelHybrid}). The resulting algorithm is summarized as Algorithm~\ref{alg:Algo1}. 
The hyperparameters of Algorithm~\ref{alg:Algo1} are the regularization coefficients $\gamma_I, \gamma_S$, the iteration limits $k_{\max}, k_{\rm proj}$, and the initial setting of $\myMat{A}^{(0)}$. These add up to the individual hyperparameters of the optimizer $\mathcal{O}_{\mathcal{L}}(\cdot)$. In our experimental study, where the number of complex \acp{adc} is equal to the number of desired sources, i.e., $P=K$, we use $\myMat{A}^{(0)} = \myMat{\Gamma}$ as our initial estimate, and set the digital filter $\myMat{B}$ via \eqref{eqn:OptimalDesB}.

\begin{algorithm}  
	\caption{Analog Combiner Setting}
	\label{alg:Algo1}
	\KwData{Fix $\myMat{A}^{(0)}$ }
	\For{$k=1,2,\ldots,k_{\max}$}{
		Update $\myMat{A}^{(k)} \leftarrow \mathcal{O}_{\mathcal{L}}\big(\myMat{A}^{(k-1)}\big)$ (see \eqref{eqn:OptConvex}) \\
		\If{$\mod(k,k_{\rm proj}) =0 $}{
		Project via $\myMat{A}^{(k)} \leftarrow \mathcal{P}_{\mathcal{A}}(\myMat{A}^{(k)})$ (see \eqref{eqn:Proj})
		}
	}
	\KwOut{Analog combiner $\myMat{A}^{(k_{\max})}$.}
\end{algorithm}

\vspace{-0.2cm}
\subsection{Discussion}
\label{subsec:Discussion}
Algorithm~\ref{alg:Algo1} allows tuning of the hybrid \ac{mimo} receiver to accurately carry out signal recovery and interferer rejection simultaneously while boosting low-power implementation. Although we do not directly optimize the power consumption, efficiency naturally follows from the usage of sparse discrete \acp{vm} and few-bit \acp{adc}. In fact, in Section~\ref{sec:Analysis}, we show that by utilizing Algorithm~\ref{alg:Algo1} to design a task-specific hybrid beamformer with low-resolution \acp{adc} and quantized \acp{vm} with sparsity, one can achieve comparable or improved \ac{mse} compared to task-agnostic hybrid receivers while consuming half the power at a $4\times$ reduced quantization rate. 

The ability to configure hybrid beamformers in a manner that is simultaneously task- and energy-aware gives rise to several challenges. The first is the fact that Algorithm~\ref{alg:Algo1} assumes accurate knowledge of the \acp{aoa} and fully-calibrated analog hardware. As in practice, one is likely to cope with mismatches, as discussed in Subsection~\ref{ssec:ModelProblem}, the design has to be extended to be robust to such inaccuracies. An additional challenge stems from the dynamic nature of wireless \ac{mimo} receivers, which implies that one should frequently reconfigure its hardware, particularly when the signal sources are mobile wireless users. The operation of Algorithm~\ref{alg:Algo1}, which involves multiple iterations, can thus be too lengthy to be repeatedly utilized in real-time. A possible way to allow the design algorithm to be carried out efficiently with a small and fixed number of iterations is by leveraging data via deep unfolding methodology \cite{monga2021algorithm, shlezinger2022model}. We leave this for future investigation.

Algorithm~\ref{alg:Algo1} assumes that the number of \acp{adc} ($P$) and their resolution ($b$) are given. Indeed, these parameters are typically fixed for a given device and are thus not considered in the optimization procedure. However, one can utilize the proposed algorithmic steps to optimize the number of bits and the number of \acp{adc}, as done in, e.g., \cite{neuhaus2020taskbased}. This can be beneficial when designing the hardware prior to its fabrication, or when employing \acp{adc} with programmable resolutions.


 
%
\section{Robust Low-Power HBF}
\label{sec:Mismatches}
In this section, we re-design our system to be robust to mismatches, as stated in Subsection~\ref{ssec:ModelProblem}. We extend Algorithm~\ref{alg:Algo1} and make it robust to inaccurate \acp{aoa} in Subsection~\ref{ssec:InacurateAoA}, and to hardware non-idealities in Subsection~\ref{ssec:HardwareMismatches}. The resulting system operation is then discussed in Subsection~\ref{ssec:DiscussionMismatch}.

\subsection{Inaccurate AoAs}
\label{ssec:InacurateAoA}

The proposed tuning of the analog combiner requires prior knowledge of the \acp{aoa} and power levels. These are used to form the correlation matrices $\CovMat{\myVec{x}} = \myMat{M}_{\theta}\CovMat{\myVec{s}}\myMat{M}_{\theta}^H + \myMat{M}_{\phi}\myMat{C}_{\myVec{v}}\myMat{M}_{\phi}^H + \sigma_w^2\myMat{I}_N$, and $\CovMat{\myVec{s}\myVec{x}} = \CovMat{\myVec{s}}\myMat{M}_{\theta}^H$. 
This information should therefore be either estimated or externally provided by a spatial sensor and may thus contain errors. 
Applying Algorithm~\ref{alg:Algo1} using the inaccurate \acp{aoa} degrades the performance. Thus, we next design a robust counterpart by formulating the design objective over the range of possible \acp{aoa} $\Theta(\myVec{\theta}, \epsilon)$, and seek to minimize the design objective for the worst-case scenario over this range. To that aim, we next detail the modification introduced to the signal recovery component of the design objective, after which we formulate the robust optimization problem and the algorithm we utilize to tackle it.

{\bf Signal Recovery:} Our goal is to reformulate the recovery error of the desired signal $\myVec{s}$ from its estimate $\hat{\myVec{s}}$ when the exact \acp{aoa} of the desired signals are unknown. For any quantized representation of $\hat{\myVec{s}}$ and any estimate of the \ac{aoa} vector $\myVec{\theta}$, the decomposition \eqref{eqn:MSEdecomposed} holds, and it is still true that only the latter term depends on the configuration of the analog pre-processing matrix $\myMat{A}$. However, the former term in \eqref{eqn:MSEdecomposed}, which represents the \ac{mmse}, depends on the \acp{aoa} of the desired signals. Consequently, if there exists any uncertainty in the desired angles, one needs to characterize the non-decomposed recovery error, $\E\{\|\myVec{s}-\hat{\myVec{s}}\|^2\}$, and can no longer consider only the distortion with respect to linear \ac{mmse} estimate $\tilde{\myVec{s}}$, $\E\{\|\tilde{\myVec{s}}-\hat{\myVec{s}}\|^2\}$, as the signal recovery performance measure. 


We again let the digital processing output an estimate of the form $\hat{\myVec{s}} = \myMat{B} \mathcal{Q}_b(\myMat{A}\myVec{x})$ for some $\myMat{B}\in\mathbb{C}^{K\times P}$ and set the \acp{adc} to have negligible overloading probability. By using the notations $\CovMat{\myVec{x}} (\myVec{\theta})$ and $\CovMat{\myVec{s}\myVec{x}}(\myVec{\theta})$ for covariance matrices in \eqref{eqn:Autocorr} to encapsulate their dependence on \acp{aoa}, we generalize the \ac{mse} expression in Lemma~\ref{lem:mse} as follows:
\begin{lemma}
\label{lem:mseAoA}
 When the \acp{adc} utilize non-subtractive dithered quantizers with vanishing overloading probability, then for any setting of the \acp{aoa} $\myVec{\theta}$, the \ac{mse} objective \eqref{eqn:MSEdecomposed} becomes
 \vspace{-0.1cm}
\begin{align}
    {\rm MSE}(\myMat{A}, \myMat{B}, \myVec{\theta}) = &{\rm Tr}\Bigg( {\CovMat{\myVec{s}}} -2 \CovMat{\myVec{s}\myVec{x}}(\myVec{\theta})\myMat{A}^H\myMat{B}^H\notag \\
    &+ \myMat{B}\myMat{A}\CovMat{\myVec{x}}(\myVec{\theta})\myMat{A}^H\myMat{B}^H\notag \\
    &+ \myMat{B}\frac{2\kappa \cdot {\rm Tr}(\myMat{A} \CovMat{\myVec{x}}(\myVec{\theta}) \myMat{A}^H)}{3b^2 \cdot P}\myMat{B}^H\Bigg).
    \label{eqn:mseAoA}
\end{align}
\end{lemma}

\begin{IEEEproof}
See Appendix~\ref{proof:lem2}.
\end{IEEEproof}

For any fixed $\myVec{\theta}$, it can be shown that the setting of $\myMat{B}$ which minimizes \eqref{eqn:mseAoA} is the one given in \eqref{eqn:OptimalDesB}, for which the \ac{mse} expression in Lemma~\ref{lem:mseAoA} coincides with the sum of the \ac{mmse} and the excess \ac{mse} of Lemma~\ref{lem:mse}. However, since in the presence of uncertainty, we do not have exact knowledge of $\myVec{\theta}$, but can only characterize a range in which it takes values, we cannot immediately substitute the \ac{mse} minimizing $\myMat{B}$ as we did in Section~\ref{sec:Non-MismatchedTS}, since each possible value of $\myVec{\theta}$ is associated with a different \ac{mse} minimizing digital filter.


Similar to Lemma~\ref{lem:mse}, Lemma~\ref{lem:mseAoA} rigorously holds when \acp{adc} are modeled as implementing non-subtractive dithered quantizers. For a broad range of input signals that do not satisfy this assumption, it also approximately holds.

{\bf Robust Optimization:} Lemma~\ref{lem:mseAoA} formulates the overall design objective for any possible $\myVec{\theta}$. After adding the regularization terms to balance the contribution of interference rejection and the power consumption of the analog circuitry, the new design objective is formulated as 
\begin{equation}
\label{eqn:objectiveFunAoA}
    \mathcal{L}(\myMat{A}, \myMat{B}, \myVec{\theta})  =  \textrm{MSE}(\myMat{A}, \myMat{B}, \myVec{\theta})\! +\! \gamma_I\textrm{IntRej}(\myMat{A}) \!+\! \gamma_S \|\myMat{A}\|_{1,1}.
\end{equation}
The robust optimization that minimizes the loss for the worst-case scenario is then given by
\begin{equation}
\label{eq:desObjAoa}
    \myMat{A}\opt, \myMat{B}\opt = \underset{\myMat{A}\in \mathcal{A}^{P \times N}, \myMat{B}\in \mySet{C}^{K \times P}}{\arg \min} \quad \underset{\myVec{\bar{\theta}} \in \Theta(\myVec{\theta}, \epsilon)}{ \max} \quad \mathcal{L}(\myMat{A}, \myMat{B}, \myVec{\bar{\theta}}).
\end{equation}

Besides the fact that the optimization problem is formulated over the possibly non-convex search space $\mathcal{A}$, the objective in \eqref{eq:desObjAoa} is also not concave with respect to the variable $\myMat{\theta}$. This makes the minimax optimization problem \eqref{eq:desObjAoa} difficult to solve. Yet, it can still be tackled with iterative solvers, even though they are not guaranteed to find the optimal solution.

{\bf Design Algorithm:} 
A candidate approach to tackle \eqref{eq:desObjAoa} is to replace the order of minimization and maximization, i.e., convert it into a maximin problem, for which the internal minimization is carried out for a given $\myVec{\theta}$, allowing to use the results of Section~\ref{sec:Non-MismatchedTS}. However, since we seek to minimize our objective, then by using the max-min inequality \cite[Ch. 5.4]{boyd2004convex}, such an approximation would result in minimizing a lower-bound on the objective \eqref{eq:desObjAoa}, and thus does not necessarily yield a suitable design in the sense of the original minimax problem. Consequently, we directly tackle \eqref{eq:desObjAoa} numerically, replacing the continuous $\Theta(\myVec{\theta}, \epsilon)$ with a set of $C$ grid points $\{\myVec{\theta}_c\}_{c=1}^C \subset \Theta(\myVec{\theta}, \epsilon)$. We thus treat the surrogate problem 
\begin{align}
\label{eq:desObjAoaSur}
    \myMat{A}\opt, \myMat{B}\opt = &\underset{\myMat{A}\in \mathcal{A}^{P \times N}, \myMat{B}\in \mySet{C}^{K \times P}}{\arg \min} \quad \underset{\gamma >0}{ \min} \quad \gamma \\
    &\text{subject to }  \mathcal{L}(\myMat{A}, \myMat{B}, \myVec{\theta}_c) \leq \gamma, \quad c\in\{1,\ldots,C\}. \notag
\end{align}

The surrogate problem in \eqref{eq:desObjAoaSur} is convex with respect to both $\myMat{A}$ and $\myMat{B}$, separately, when the analog combiner is unconstrained, i.e., $\mathcal{A}=\mathbb{C}$. Consequently, when $\mathcal{A}=\mathbb{C}$, \eqref{eq:desObjAoaSur} can be tackled using alternating optimization over both $\myMat{A}$ and $\myMat{B}$. We convert the constrained \eqref{eq:desObjAoaSur} into an unconstrained optimization by formulating the logarithmic barrier function~\cite{Waltz2006_InteriorPoint} 
%
\begin{align}
    \textcolor{red}{\mathcal{F}}(\myMat{A}, \gamma, \myMat{B}) = & \gamma - \mu_h \sum_{c = 1}^C {\rm ln}(\gamma - \textrm{MSE}(\myMat{A}, \myMat{B}, \myVec{\theta}_c))\notag \\ 
     &+ \gamma_I\textrm{IntRej}(\myMat{A}) + \gamma_S \|\myMat{A}\|_{1,1},\label{eqn:RobOptA}
\end{align}
where $\mu_h>0$ is the barrier hyperparameter.

We next let $\bar{\mathcal{O}}_{\mathcal{F}_{\myMat{A}}}(\cdot)$ and $\bar{\mathcal{O}}_{\mathcal{F}_{\myMat{B}}}(\cdot)$ be iterative optimizers of \eqref{eqn:RobOptA} with respect to $(\myMat{A}, \gamma)$ and to $(\myMat{B}, \gamma)$ respectively. Similarly to the approach adopted in Subsection~\ref{subsec:Design}, this can be achieved by defining the task term as $\tilde{\mathcal{F}}(\myMat{A}, \gamma, \myMat{B}) := \gamma - \mu_h \sum_{c = 1}^C {\rm ln}(\gamma - \textrm{MSE}(\myMat{A}, \myMat{B}, \myVec{\theta}_c)) + \gamma_I\textrm{IntRej}(\myMat{A})$. We can now set $\bar{\mathcal{O}}_{\mathcal{F}_{\myMat{A}}}(\cdot)$ to be proximal gradient steps of the form
\begin{equation}
\label{eqn:OptA}
    \bar{\mathcal{O}}_{\mathcal{F}_{\myMat{A}}}(\myMat{A}, \gamma, \myMat{B}) \!=\! \mathcal{T}_{\gamma_S}\left\{(\myMat{A}, \gamma) \!-\! \mu_{\myMat{A}}\nabla_{\myMat{A}, \gamma} \tilde{\mathcal{F}}(\myMat{A}, \gamma, \myMat{B})\right\},
\end{equation}
where $\mathcal{T}$ denotes element-wise complex soft-thresholding (as in \eqref{eqn:OptConvex}). Similarly,  $\bar{\mathcal{O}}_{\mathcal{F}_{\myMat{A}}}(\cdot)$ is a gradient step 
\begin{equation}
\label{eqn:OptB}
    \bar{\mathcal{O}}_{\mathcal{F}_{\myMat{B}}}(\myMat{A},\gamma, \myMat{B}) = (\myMat{B}, \gamma) - \mu_{\myMat{B}}\nabla_{\myMat{B}, \gamma} \tilde{\mathcal{F}}(\myMat{A}, \gamma, \myMat{B}).
\end{equation}
In \eqref{eqn:OptA}-\eqref{eqn:OptB}, the hyperparameters $\mu_{\myMat{A}}, \mu_{\myMat{B}}$ are the step-sizes. As in \eqref{eqn:OptConvex}, the differentiability of $\tilde{\mathcal{F}}(\myMat{A}, \gamma, \myMat{B})$ allows computing its gradients using automatic differentiation tools.


The optimization for tackling \eqref{eqn:RobOptA} can be based on, e.g., alternating between a fixed number of iterations for setting $\myMat{A}$ using $\bar{\mathcal{O}}_{\mathcal{F}_{\myMat{A}}}(\cdot)$, followed by a fixed number of iterations for updating $\myMat{B}$ via $\bar{\mathcal{O}}_{\mathcal{F}_{\myMat{B}}}(\cdot)$, as we do in our numerical study in Section~\ref{sec:Analysis}.
%
We follow the design strategy proposed in Subsection~\ref{subsec:Design}, and augment its iterations with periodic projections onto the discrete set $\mathcal{A}$. To facilitate convergence to a suitable setting of $\myMat{A}$ and $\myMat{B}$, we set their initial values to be the $\myMat{A}^{(0)} = \myMat{\Gamma}(\myVec{\theta})$, and the digital filter $\myMat{B}^{(0)}$ via \eqref{eqn:OptimalDesB}. The resulting algorithm is summarized as Algorithm~\ref{alg:Algo2}.

\begin{algorithm}  
	\caption{Robust Beamforming Design}
	\label{alg:Algo2}
	\KwData{Set $\myMat{A}^{(0)}$, $\myMat{B}^{(0)}$}
	\For{$k=1,2,\ldots,k_{\max}$}{
	    Set $\bar{\myMat{A}}^{(0)} \leftarrow \myMat{A}^{(k-1)}$, $\bar{\myMat{B}}^{(0)} \leftarrow \myMat{B}^{(k-1)}$\;
	    Set $\gamma  \leftarrow  \max_c \mathcal{L}(\bar{\myMat{A}}^{(0)}, \bar{\myMat{B}}^{(0)}, \myVec{\theta}_c)$ via \eqref{eqn:objectiveFunAoA}\;
	    \For{$l=1,2,\ldots,l_{\max}$}{
	        Update $\bar{\myMat{A}}^{(l)}, \gamma \leftarrow  \bar{\mathcal{O}}_{\mathcal{F}_{\myMat{A}}}(\bar{\myMat{A}}^{(l-1)}, \gamma, {\myMat{B}}^{(k-1)})$\;
	    }
		Update $\myMat{A}^{(k)} \leftarrow \bar{\myMat{A}}^{(l_{\max})}$\;
		\If{$\mod(k,k_{\rm proj}) =0 $}{
		Project via $\myMat{A}^{(k)} \leftarrow \mathcal{P}_{\mathcal{A}}(\myMat{A}^{(k)})$
		}
	    \For{$l=1,2,\ldots,l_{\max}$}{
	        Update $\bar{\myMat{B}}^{(l)}, \gamma \leftarrow  \bar{\mathcal{O}}_{\mathcal{F}_{\myMat{B}}}({\myMat{A}}^{(k)}, \gamma, \bar{\myMat{B}}^{(l-1)})$\;
	    }
	    Update  $\myMat{B}^{(k)} \leftarrow \bar{\myMat{B}}^{(l_{\max})}$\;
	}
	\KwOut{Analog combiner $\myMat{A}^{(k_{\max})}$ and digital processing matrix $\myMat{B}^{(k_{\max})}$.}
\end{algorithm}

%
\subsection{Hardware Non-Idealities}
\label{ssec:HardwareMismatches}
As shown in Fig.~\ref{fig:VM_architecture}, \acp{vm} are implemented with a quadrature all-pass filter and two variable gain amplifiers (\acp{vga}). Quadrature all-pass filter generates the required I/Q components, while \acp{vga} determine the set of complex phase/gain states that the \ac{vm} can be configured to exhibit, i.e., the set $\mathcal{A}$. In the system described in Section~\ref{sec:Non-MismatchedTS}, the values in $\mathcal{A}$ are assumed to be precisely known and can be ideally configured. However, when \acp{vm} are employed in practice, various hardware non-idealities such as gain and phase mismatches at I/Q components or at the output of \ac{vm}, make it impossible to realize the exact desired complex gains. 
These mismatches can be measured and are usually calibrated in the digital domain, which is less challenging than designing analog hardware with stricter design performance specifications. However, our need to operate low-resolution \acp{adc} requires tightly coupled high-precision analog pre-processing and digital processing and necessitates calibrating the analog pre-processing. In the following, we integrate analog hardware calibration into our system design by incorporating phase and gain mismatches into the algorithm. To this end, we first propose a hardware-compliant model for the mismatches. The effect of mismatches is captured by scaling or rotation of the feasible mapping between \acp{vm} and the set $\mathcal{A}$. 

\begin{figure}
    \centering
    \includegraphics[width=\columnwidth]{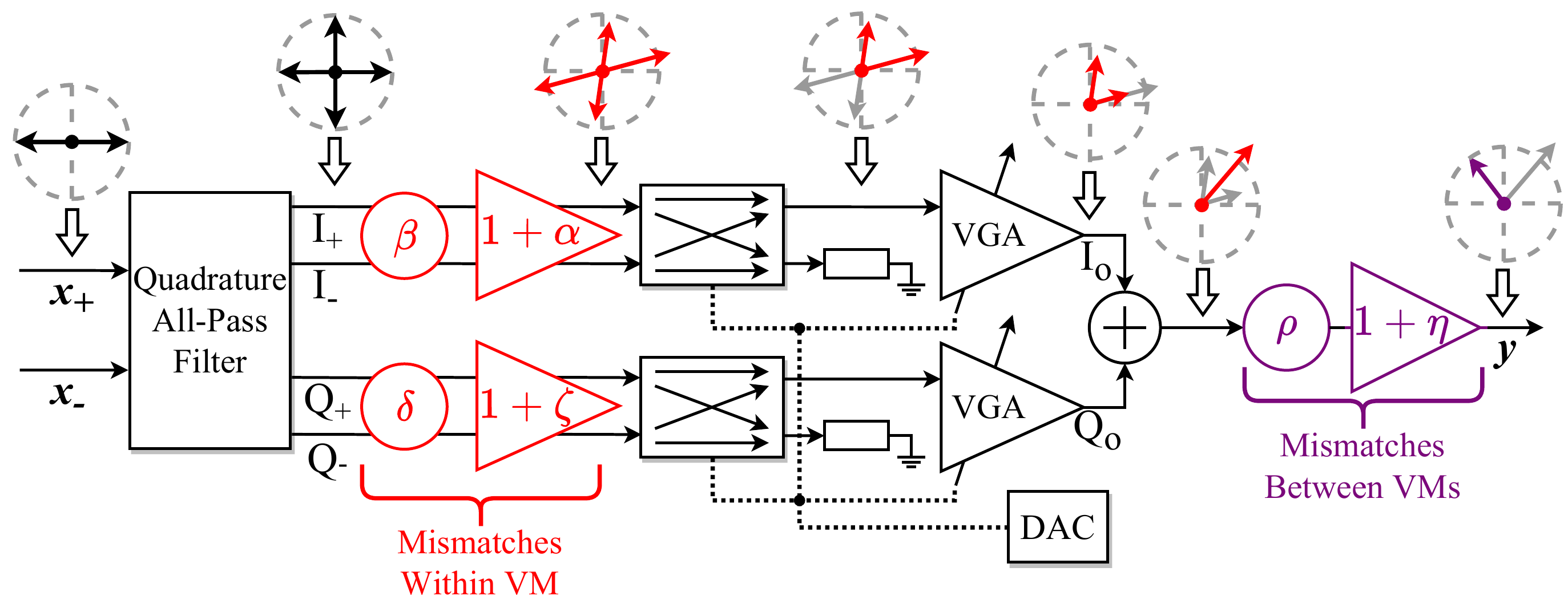}
    \caption{\ac{vm} architecture model with I/Q mismatches.}
    \label{fig:TB_MIMO_VM_mismatch}
\end{figure}

\textbf{Hardware Compliant Model:} The mismatched hardware model is illustrated in Fig.~\ref{fig:TB_MIMO_VM_mismatch}. We denote gain and phase mismatches as $(\alpha_{p, n}, \beta_{p, n})$ and $(\zeta_{p, n}, \delta_{p, n})$ corresponding to the I/Q paths within the \ac{vm} $[\boldsymbol{A}]_{p, n}$, respectively. Similarly, mismatches between different \acp{vm} stemming from routing delay and attenuation in practical hardware can be observed at the output paths of the \acp{vm}. Mismatches between different\acp{vm} are modeled at the output of 
Fig.~\ref{fig:TB_MIMO_VM_mismatch} by letting $(\eta_{p, n}, \rho_{p, n})$  be gain and phase mismatch corresponding to the output path of the $[\boldsymbol{A}]_{p, n}$ entry, respectively. Under the mismatches in the \ac{vm} corresponding to the $[\boldsymbol{A}]_{p, n}$ entry, the corrupted discrete set $\mathcal{\Bar{A}}_{p,n}$, which the \ac{vm} can realize, is given as:
\begin{align}
\label{eq:setA_HWMismatchVM}
     &\mathcal{\Bar{A}}_{p,n} = \big \{(1 + \alpha_{p,n})(1 + \eta_{p,n}) e^{j(\beta_{p,n} + \rho_{p,n})} {\rm Re}\{a\} \notag  \\ &    + (1 + \zeta_{p,n})(1 + \eta_{p,n}) j e^{j(\delta_{p,n} + \rho_{p,n})} {\rm Im}\{a\} \mid a \in \mathcal{A} \big \}.
\end{align}

\begin{figure}
\centering
  \subfloat[]{
       \includegraphics[width=0.45\linewidth]{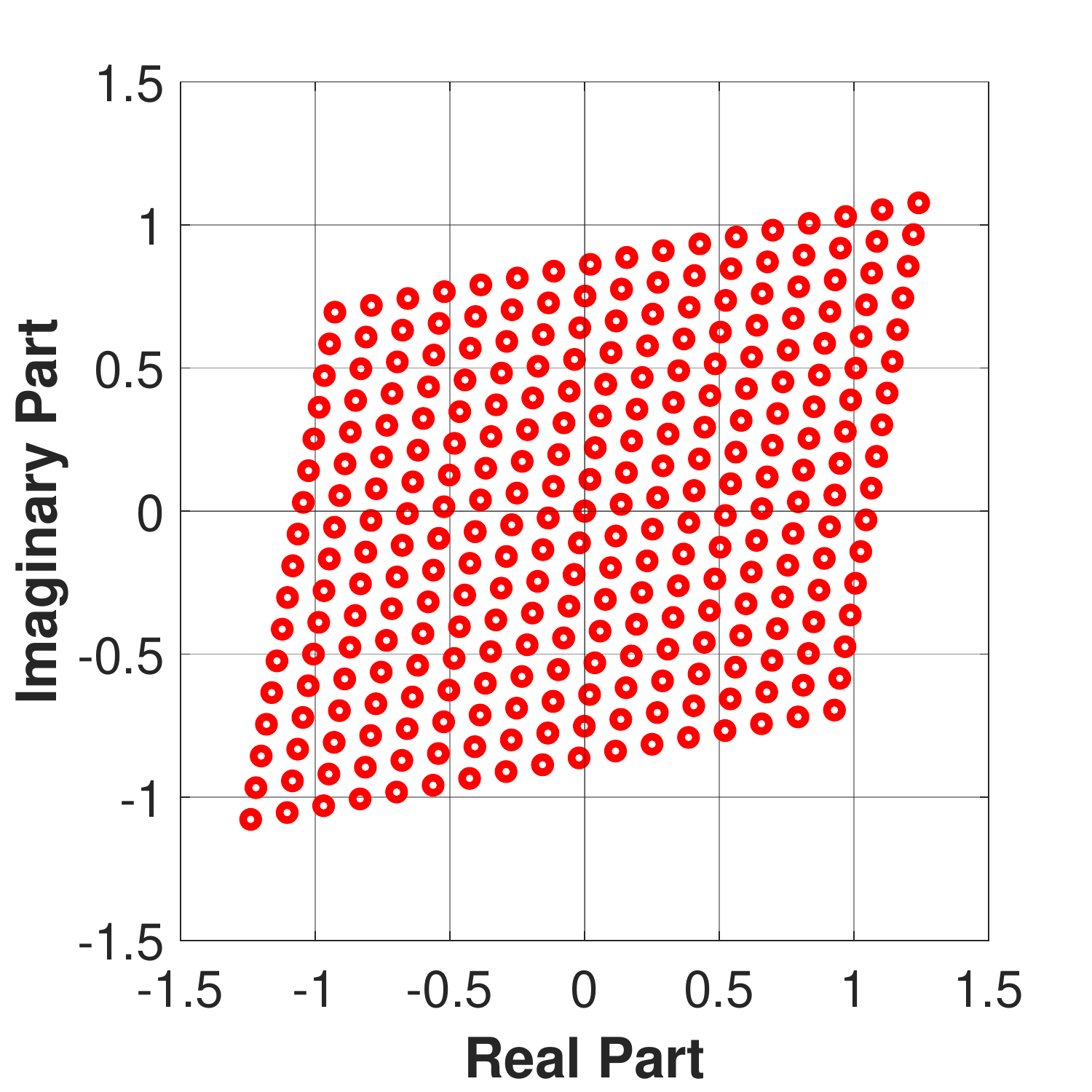}}
    \hfill
  \subfloat[]{
        \includegraphics[width=0.45\linewidth]{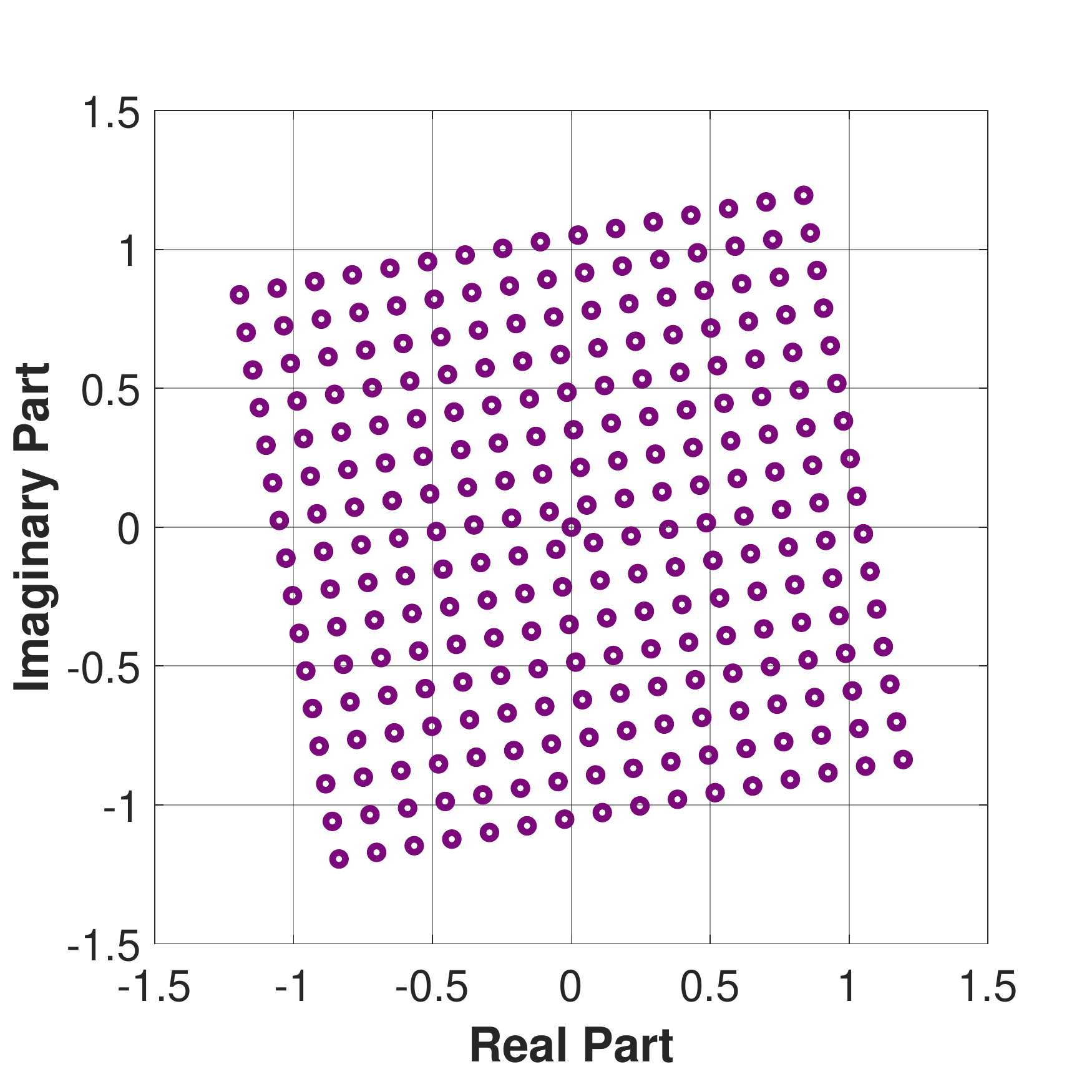}}
  \caption{4-bit \ac{vm} constellation under mismatches modeled by \eqref{eq:setA_HWMismatchVM} (a) $(\alpha_{p,n}, \beta_{p,n}) \!=\! (-\zeta_{p,n}, -\delta_{p,n}) \!=\! (0.1, 10^\circ)$; (b) $(\eta_{p,n}, \rho_{p,n}) \!=\! (0.1, 10^\circ).$} 
  \label{fig:VM_2_mismatch} 
\end{figure}

To construct the corrupted set $\mathcal{\Bar{A}}_{p,n}$, we need to measure the hardware mismatches. This is generally done by deactivating all except one \ac{vm} and measuring the output. Hence, phase and gain mismatches are assumed known. The ideal case without any mismatch corresponds to $\alpha_{p,n} = \zeta_{p,n} = \eta_{p,n} = 0$ and $\beta_{p,n} = \delta_{p,n} = \rho_{p,n} = 0^\circ$ for all $(p,n)$,
i.e. for every \ac{vm}. Examples of mismatched normalized complex constellation diagrams are illustrated in Fig.~\ref{fig:VM_2_mismatch}.

\textbf{Co-optimized Design:} 
We next incorporate the proposed hardware-compliant model into our design, allowing it to cope with these mismatches. Compensating for hardware mismatches requires the knowledge of the gain and phase mismatches, which can be measured from the real hardware. After these mismatches are measured, we update the optimization problem in \eqref{eqn:OptProblem} to project onto the new set $\mathcal{A}$
\begin{equation}
\label{eqn:OptProblemIQImbalance}
    \myMat{A}\opt = \underset{[\boldsymbol{A}]_{p, n}\in \mathcal{\Bar{A}}_{p,n}}{\arg \min} \mathcal{L}(\myMat{A}).
\end{equation}
The optimization problem \eqref{eqn:OptProblemIQImbalance} has the same structure with \eqref{eqn:ObjectiveA}, while having a different projection set. After defining the new projection set as $\bar{\mathcal{A}} =\{\myMat{A}: [\boldsymbol{A}]_{p, n}\in \mathcal{\Bar{A}}_{p,n}\}$, problem \eqref{eqn:OptProblemIQImbalance} is tackled by changing the projection step $\mathcal{P}_{\mathcal{A}}(\myMat{A})$ with $\mathcal{P}_{\bar{\mathcal{A}}}(\myMat{A})$ in Algorithm~\ref{alg:Algo1}. While projection is still done element-wise, here $\mathcal{P}_{\bar{\mathcal{A}}}(\myMat{A})$ cannot be obtained by separately quantizing the I/Q components, as in \eqref{eqn:Proj}, and must be done directly, i.e., 
\begin{equation*}
       \big[\mathcal{P}_{\bar{\mathcal{A}}}(\myMat{A}) \big]_{p,n} = \mathop{\arg\min}\limits_{a \in \bar{\mathcal{A}}} \big\|[\myMat{A}]_{p,n}-a\big\|. 
\end{equation*}

\subsection{Discussion}
\label{ssec:DiscussionMismatch} 

Problem \eqref{eq:desObjAoa} represents the robust design as minimax optimization, in which the set over which maximization is carried out, $\Theta(\myVec{\theta}, \epsilon)$, is continuous. While there are minimax optimization algorithms designed to operate over continuous domains \cite{thekumparampil2019efficient}, for simplicity, we adopt a method geared towards finite maximization sets. In our numerical study, we observed that the performance of the system is not sensitive to the choice of the number of grid points, i.e. $C$.

The design of a robust system described in Subsection~\ref{ssec:InacurateAoA} accounts for only the uncertainties in the \acp{aoa} of the desired signals. Nonetheless, one can follow the same algorithmic approach to make the system robust against other uncertainties such as the \acp{aoa} of interferers and power levels of the signals. This study is left for future research.

The design proposed in Subsection~\ref{ssec:HardwareMismatches} facilitates coping with hardware mismatches by measuring and incorporating them into the design procedure. Our algorithm relies on our proposed modeling of non-idealities such as amplitude-dependent phase shift in the \acp{vga} \cite{Mayer2011_OtherMismatch}. Similarly, other \ac{vm} implementations can give rise to different types of mismatches. Our design naturally extends to other forms of mismatches, as once the complex set $\mathcal{A}$ and its mismatched counterpart $\mathcal{\bar{A}}$ are correctly defined, it can be incorporated into Algorithms~\ref{alg:Algo1}-\ref{alg:Algo2}.


\section{Numerical Evaluations}
\label{sec:Analysis}
In this section, we evaluate  the proposed hybrid MIMO receiver. We present the simulation setup in Subsection~\ref{ssec:SimSetup}. Next, in Section~\ref{ssec:MSE}, we characterize the MSE performance in signal recovery with and without mismatches and show robustness to mismatches. We demonstrate the system's ability to reject spatial interferers in Subsection~\ref{ssec:Beampatterns}, and provide power consumption estimates in Subsection~\ref{ssec:PowerConsumption}.

\subsection{Simulation Setup}
\label{ssec:SimSetup}
\begin{figure}
    \centering
    \includegraphics[width=\columnwidth]{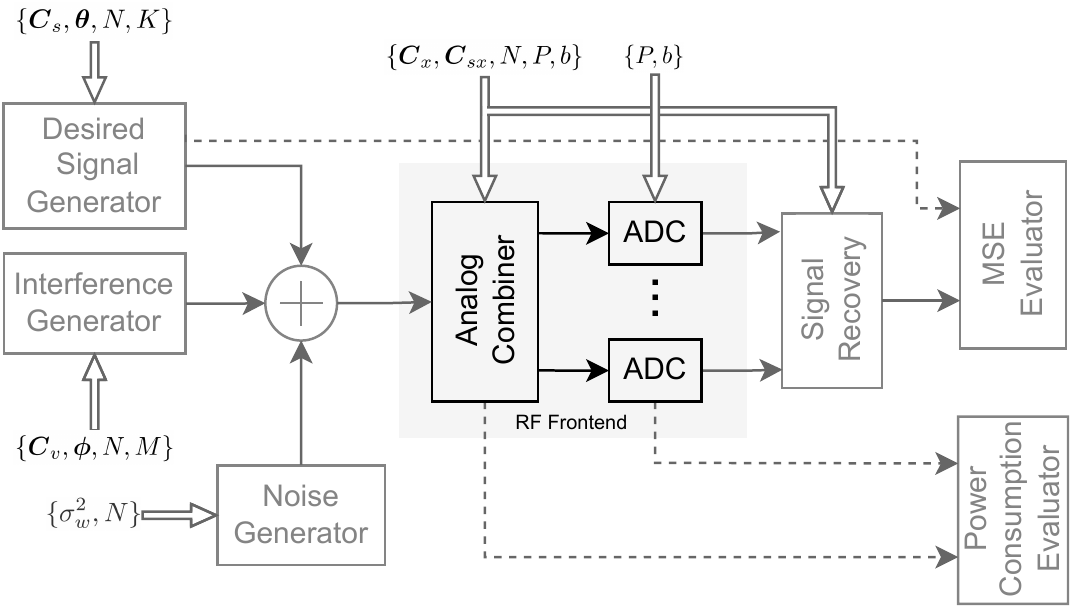}
    \caption{Task-specific hybrid MIMO receiver system testbench.}
    \label{fig:Testbench}
\end{figure}

To evaluate the task-agnostic MIMO receiver and estimate power savings, we utilize the simulation testbench illustrated in Fig.~\ref{fig:Testbench}. 
The hardware system features an RF front end as an analog combiner and digital signal processing back end for signal recovery, as described in Subsection~\ref{ssec:ModelHybrid}. The $N$-element input  $\myVec{x}$ is fed into the analog combiner $\myMat{A}$ at RF frequencies, down-converted to a low baseband frequency, and provided to the low-quantization rate ADCs as I/Q signals. In the digital domain, the filter $\myMat{B}$ is applied and signal recovery is performed. Recovery performance is characterized using MSE. Power savings of the system is evaluated over the total power consumption of the analog combiner and ADCs.

We simulate $K=2$ desired signals at angles $\theta_1= \frac{\pi}{8}, \theta_2= -\frac{\pi}{4}$ with variances $1.5$ and $0.5$, respectively. We model $M=2$ unwanted interferers at angles $\phi_1= -\frac{\pi}{18}, \phi_2= \frac{\pi}{3}$ with a variance of $5$ for both sources. Consequently, the interferers are several times stronger than the desired signals. We define the \ac{snr} as the ratio between the average power level of desired signals to the power level of noise, $\sigma_w^2$. 
We consider a hybrid \ac{mimo} receiver with $N=8$ half-wavelength spaced antenna elements and $P=2$ RF chains, and $2\cdot P = 4$ \acp{adc} ({\em TS Hybrid}).
For comparison purposes, we simulate a fully-digital \ac{mimo} receiver with infinite resolution quantization ({\em No Quant.}), whose \ac{mse} constitutes a lower bound on the achievable performance. We also simulate two \ac{mimo} receivers operating with quantization constraints: a fully-digital \ac{mimo} receiver ({\em Fully Digital}) and a task-agnostic hybrid \ac{mimo} system ({\em TA Hybrid}), where the analog combiner is tuned to beamform towards the \acp{aoa} $\myVec{\theta}$.

\subsection{MSE Performance}
\label{ssec:MSE}

\textbf{Non-mismatched Performance:} We first evaluate the signal recovery \ac{mse} performance achieved using Algorithm~\ref{alg:Algo1} for various \ac{vm} resolutions and sparsity levels of the analog matrix $\myMat{A}$, without any non-idealities. 
The coefficients of the analog-combiner matrix $\myMat{A}$ are computed via Algorithm~\ref{alg:Algo1}. 

\begin{figure}
    \centering
    \includegraphics[width=\figwidth]{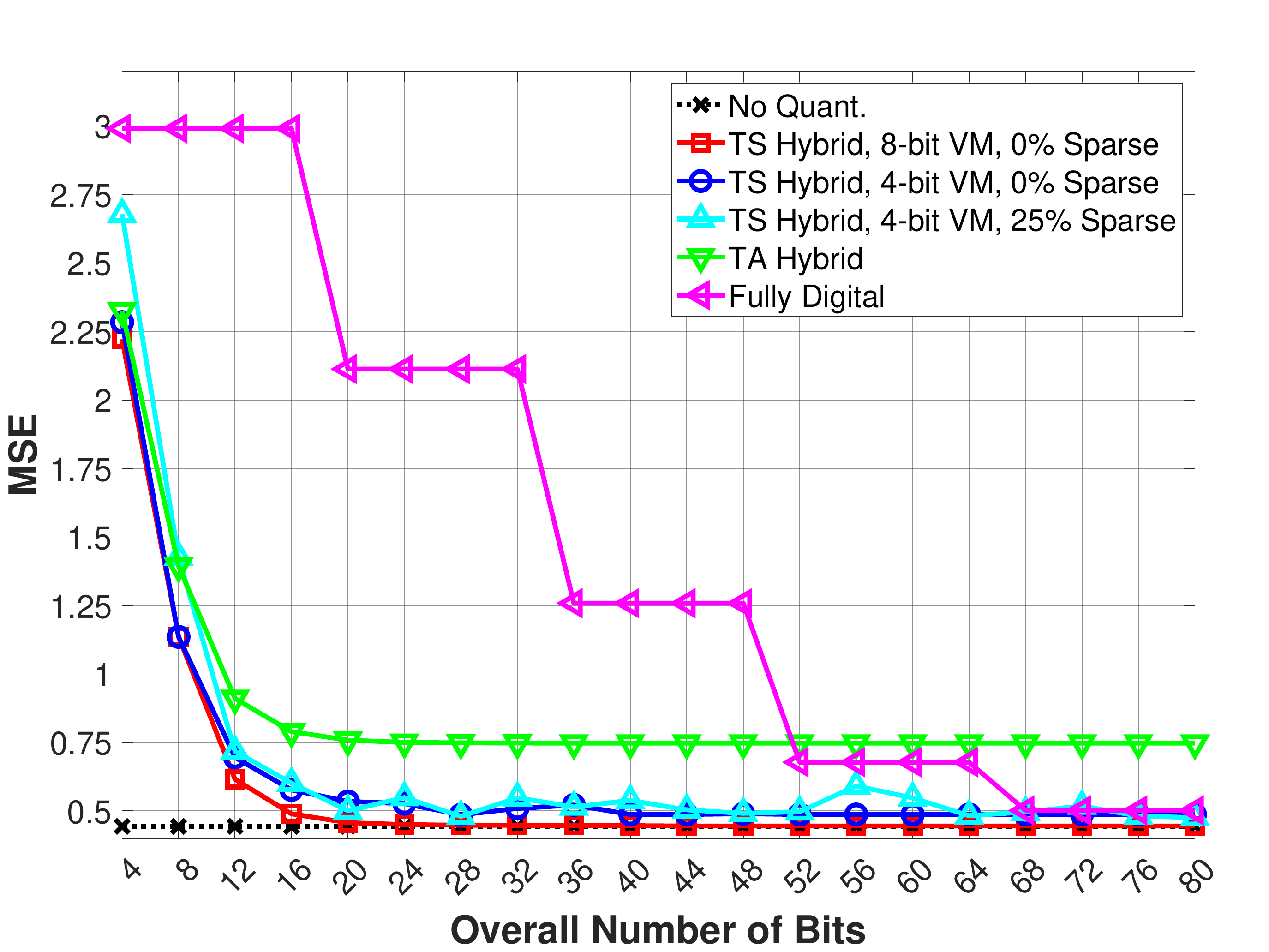}
    \caption{MSE  vs. the total number of quantization bits (SNR $= 0$ dB).}
    \label{fig:TB_MIMO_sweepBitsUnmismatched}
\end{figure}

The MSE in recovering $\myVec{s}$ versus the overall number of bits, i.e., $2 \cdot P \lceil \log_2 b\rceil$, is depicted in Fig.~\ref{fig:TB_MIMO_sweepBitsUnmismatched}. We show the numerical simulation results for 0$\%$ sparse operation with 8-bit \acp{vm}, and for 0$\%$ and 25$\%$ sparse $\myMat{A}$ with \acp{vm} quantized with relatively low resolution, e.g., 4-bit resolution. The SNR is set to 0 dB. We observe in Fig.~\ref{fig:TB_MIMO_sweepBitsUnmismatched} that by utilizing Algorithm~\ref{alg:Algo1}, one can design a task-specific hybrid \ac{mimo} receiver using low-quantization rate ADCs to approach the \ac{mse} performance achieved without any quantization, while using low-resolution \acp{vm}, e.g., merely $2^4=16$ different settings for 4 bits, and deactivating $25\%$ of VMs for sparsity to reduce power.
The task-agnostic fully-digital \ac{mimo} receiver achieves substantially worse \ac{mse} at a comparable total ADC bit budget. For the same targeted \ac{mse} , more than $4\times$ quantization rate reduction is observed with the proposed method. The conventional hybrid \ac{mimo} receiver, which only beamforms towards the desired angles, also demonstrates worse recovery \ac{mse}. Thus, the proposed method shows $1.5\times$ lower MSE at the same quantization rate of 16 bits, as shown in Fig.~\ref{fig:TB_MIMO_sweepBitsUnmismatched}.

\begin{figure}
\centering
    \includegraphics[width=\figwidth]{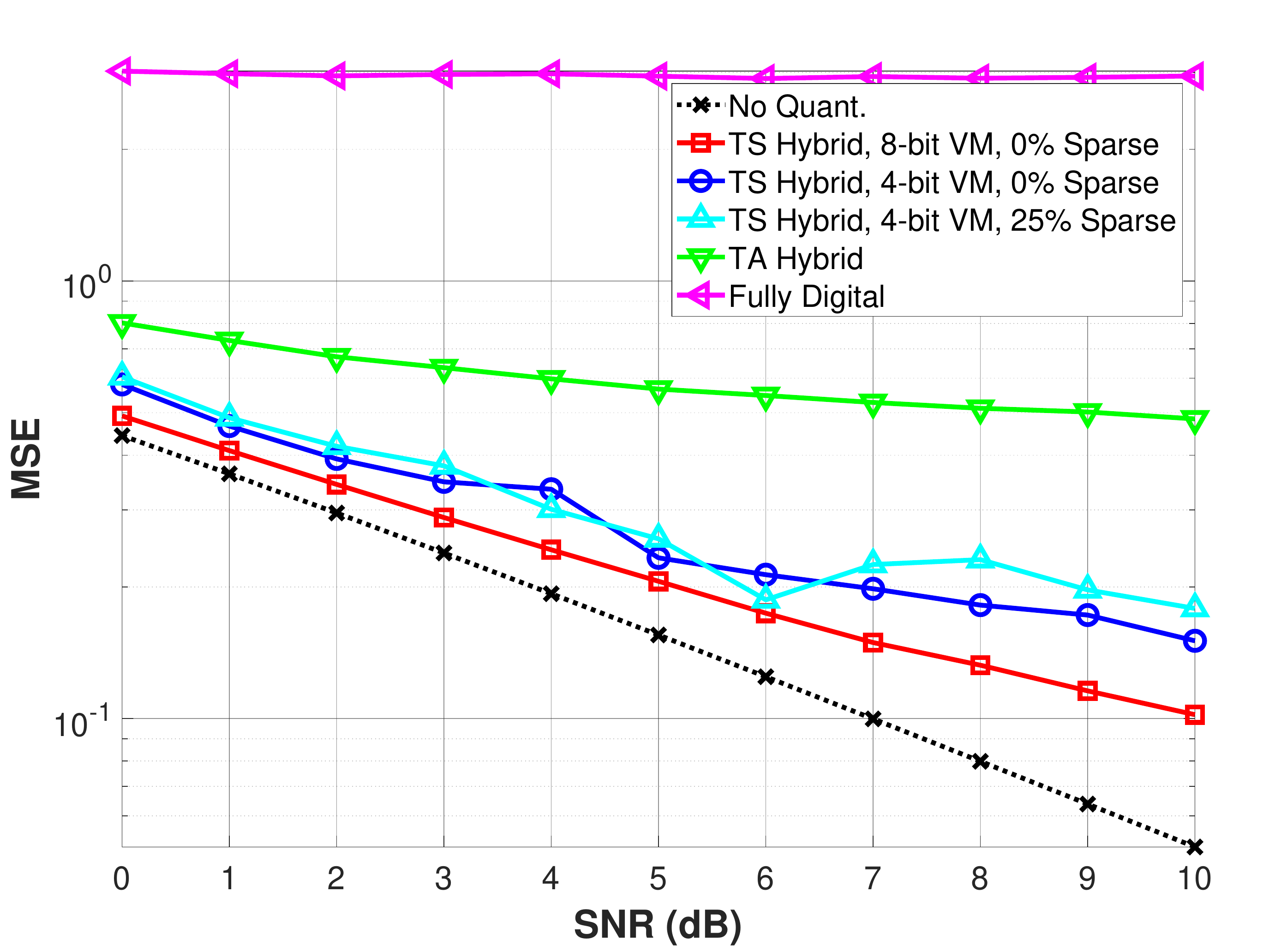}
    \caption{MSE  vs. SNR (Overall number of bits = 16).}
    \label{fig:TB_MIMO_sweepSNRUnmismatched}
\end{figure}

The effect of the noise level on the recovery performance of the system is studied next. The MSE in recovering the desired signals versus the SNR is depicted in Fig.~\ref{fig:TB_MIMO_sweepSNRUnmismatched}. The power level of the noise is swept whereas the power level of desired signals and interferences are unchanged to generate different SNRs ranging from $0$dB to $10$dB. The overall number of bits is 16, which is found to be a relatively low quantization rate where task-specific recovery via Algorithm~\ref{alg:Algo1} approaches to the MSE achieved without quantization.
Here, the fully digital task-agnostic MIMO receiver achieves the highest \ac{mse}, and does not approach the performance floor at any SNR level. The conventional hybrid MIMO receiver achieves better MSE than the quantized fully-digital system, however, it is not able to deliver the same recovery \ac{mse} as the proposed algorithm. The MSE performance of the proposed method with different resolutions and sparsity levels achieves recovery performance close to the performance floor at every level of SNR, significantly outperforming the benchmark systems.

\textbf{Performance under Inaccurate \acp{aoa}:} We next study the recovery performance of the robust and non-robust designs when there exists an uncertainty in the \acp{aoa}. 
We evaluate recovery performance for 0$\%$ sparse $\myMat{A}$ with \acp{vm} quantized with 4-bit resolution. Error margin parameter $\epsilon$ is swept from $0^\circ$ to $10^\circ$, and random \ac{aoa} realizations within the interval $\Theta(\myVec{\theta}, \epsilon)$ are generated for each $\epsilon$ value. Subsequently, the MSE in the recovery of desired signals for all realizations is calculated for analog combiner $\myMat{A}$ which is configured by non-robust Algorithm~\ref{alg:Algo1} and robust Algorithm~\ref{alg:Algo2}. We show the worst recovery \ac{mse} among all the \ac{aoa} realizations. Finally, we evaluate the recovery \ac{mse} when the \acp{aoa} are accurate but the system is configured with the  robust Algorithm~\ref{alg:Algo2}.

\begin{figure}
    \centering
    \includegraphics[width=\figwidth]{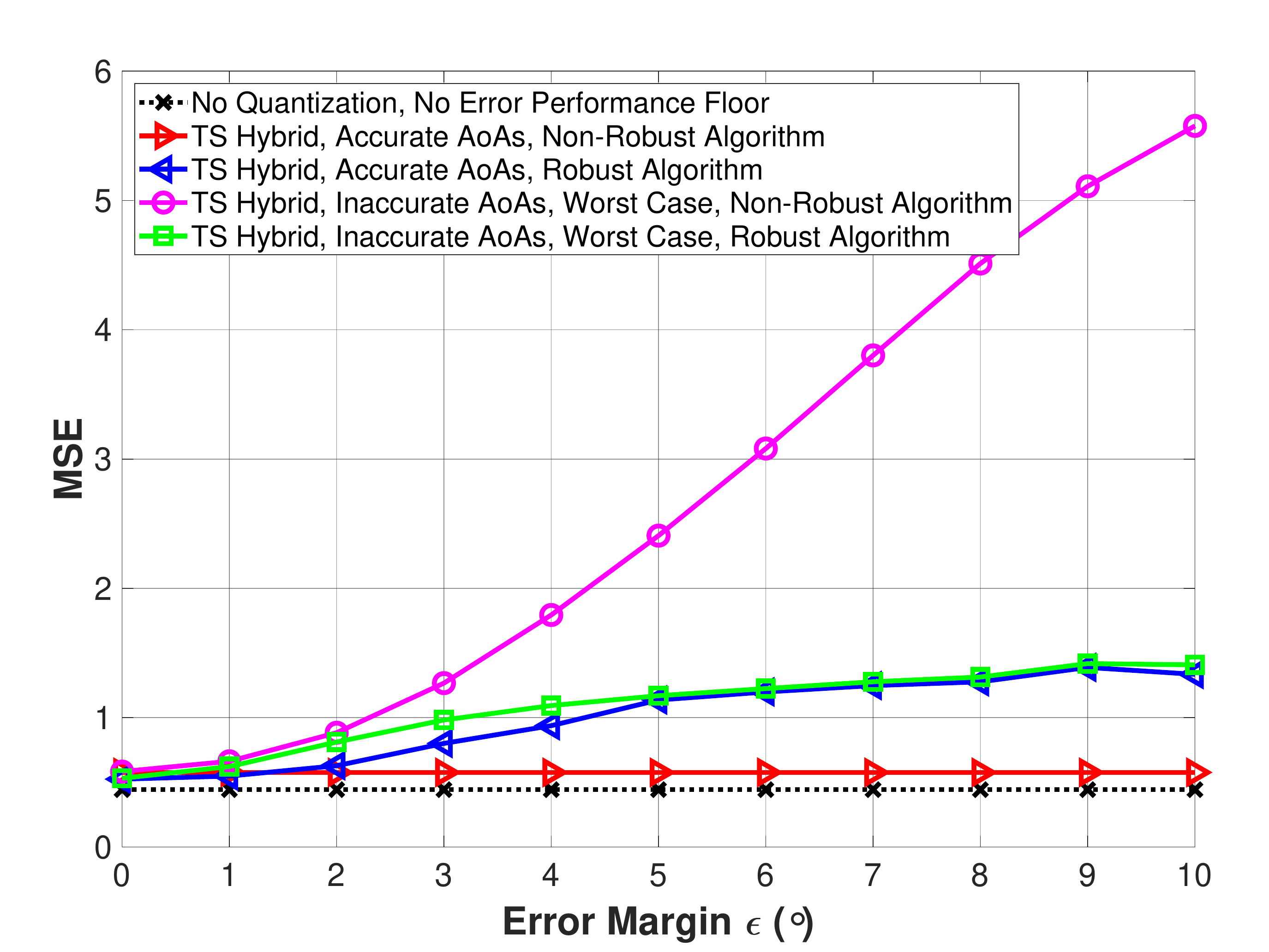}
    \caption{MSE  vs. error margin $\epsilon$ (SNR $= 0$ dB, overall number of bits = 16, TS Hybrid, 4-bit VM, 0$\%$ sparse).}
    \label{fig:TB_MIMO_AoA_sweepErrorMargin}
\end{figure}

The MSE in recovering $\myVec{s}$ versus the error margin $\epsilon$ is depicted in Fig.~\ref{fig:TB_MIMO_AoA_sweepErrorMargin}. We set the SNR level to 0 dB and the overall number of bits to 16 bits. The system configured via the robust algorithm achieves better recovery performance for the worst-case scenario than a non-robust algorithm for all degrees of uncertainties. For error margins higher than $2^\circ$ degrees, the robust system notably outperforms the non-robust system in terms of signal recovery. However, because of the fact that the robust algorithm minimizes the recovery error for the worst-case scenario over the range of possible angles, it is observed that if the system is configured with the robust algorithm even though the \acp{aoa} are accurate, worse \ac{mse} than the non-robust algorithm is obtained. Hence, one should use Algorithm~\ref{alg:Algo2} only if the \acp{aoa} are doubted to be accurate.

\begin{figure}
    \centering
    \includegraphics[width=\figwidth]{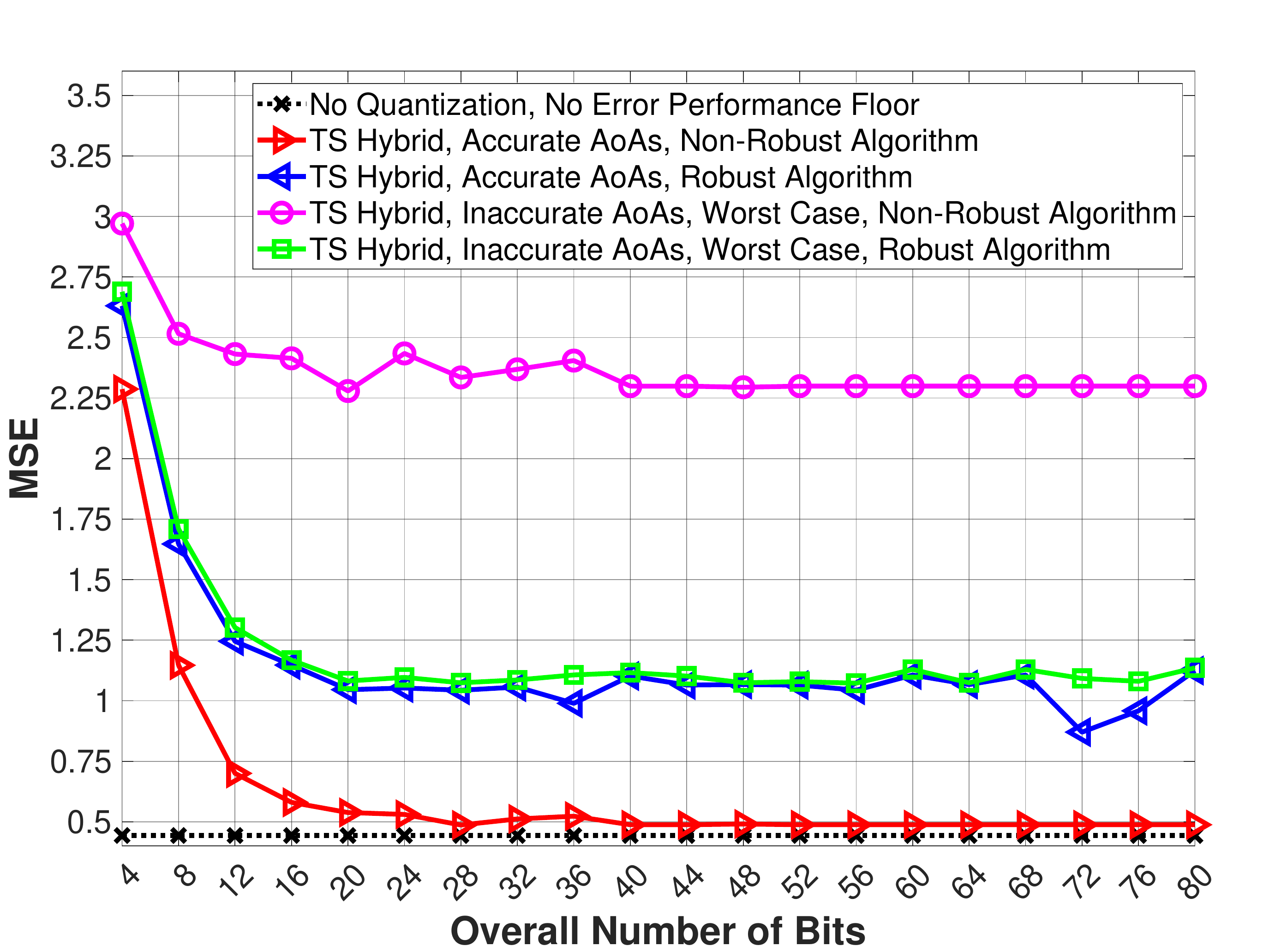}
    \caption{MSE  vs. overall number of bits (SNR $= 0$ dB, error margin $\epsilon = 5^\circ$, 4-bit VM, 0$\%$ sparse).} \label{fig:TB_MIMO_AoA_sweepBits}
\end{figure}

Next, we show how the MSE in recovering $\myVec{s}$ is affected by the overall number of bits when the \acp{aoa} are known up to some degree. We set the error margin to $\epsilon = 5^\circ$ and the SNR level to 0 dB. The results, reported in Fig.~\ref{fig:TB_MIMO_AoA_sweepBits}, demonstrate that the hybrid \ac{mimo} system configured via a non-robust Algorithm~\ref{alg:Algo1} receiver achieves worse recovery performance at a comparable total ADC bit budget. The proposed robust Algorithm~\ref{alg:Algo2} shows $2.1\times$ lower MSE than the non-robust Algorithm~\ref{alg:Algo1} at the quantization rate of $\geq$16 bits for the worst-case scenario in the possible range of \acp{aoa}. However, it shows $2.1\times$ higher MSE than non-robust Algorithm~\ref{alg:Algo1} at the same quantization rate of $\geq$16 bits when the \acp{aoa} are accurate.

\textbf{Performance under Hardware Non-Idealities:} We next consider the receiver operation under phase and gain mismatches in the \acp{vm}. The SNR and the overall number of bits are set to 0 dB and 16 bits, respectively. 
We evaluate the recovery performance for 25$\%$ sparse combiner with 4-bit \acp{vm}, comparing $\myMat{A}$ configured via Algorithm~\ref{alg:Algo1}, which does not take into account the mismatches, and its co-optimized version that is described in Subsection~\ref{ssec:HardwareMismatches}. We present task-agnostic hybrid MIMO system performance under phase and gains mismatches as benchmarks, as well as show the system's performance without any mismatches. 
In practice, the total amount of such mismatches is random, but there are fixed errors resulting from manufacturing processes and long-term aging effects. Consequently, in \cite{deepak2019impairments, bakr2009impairments}, mismatches are modeled using random variables. Hence, to calculate the MSE in recovery under a practical scenario of gain and phase mismatches, we generate random realizations of phase and gain mismatch pairs, uniformly distributed around a given mean, with variations of $\pm 10\%$. In the co-optimized algorithm, we tune the analog combiner via the mean of the mismatches. The MSE performance of the system is evaluated by average recovery error among all realizations. 

\begin{figure}
    \centering
    \includegraphics[width=\figwidth]{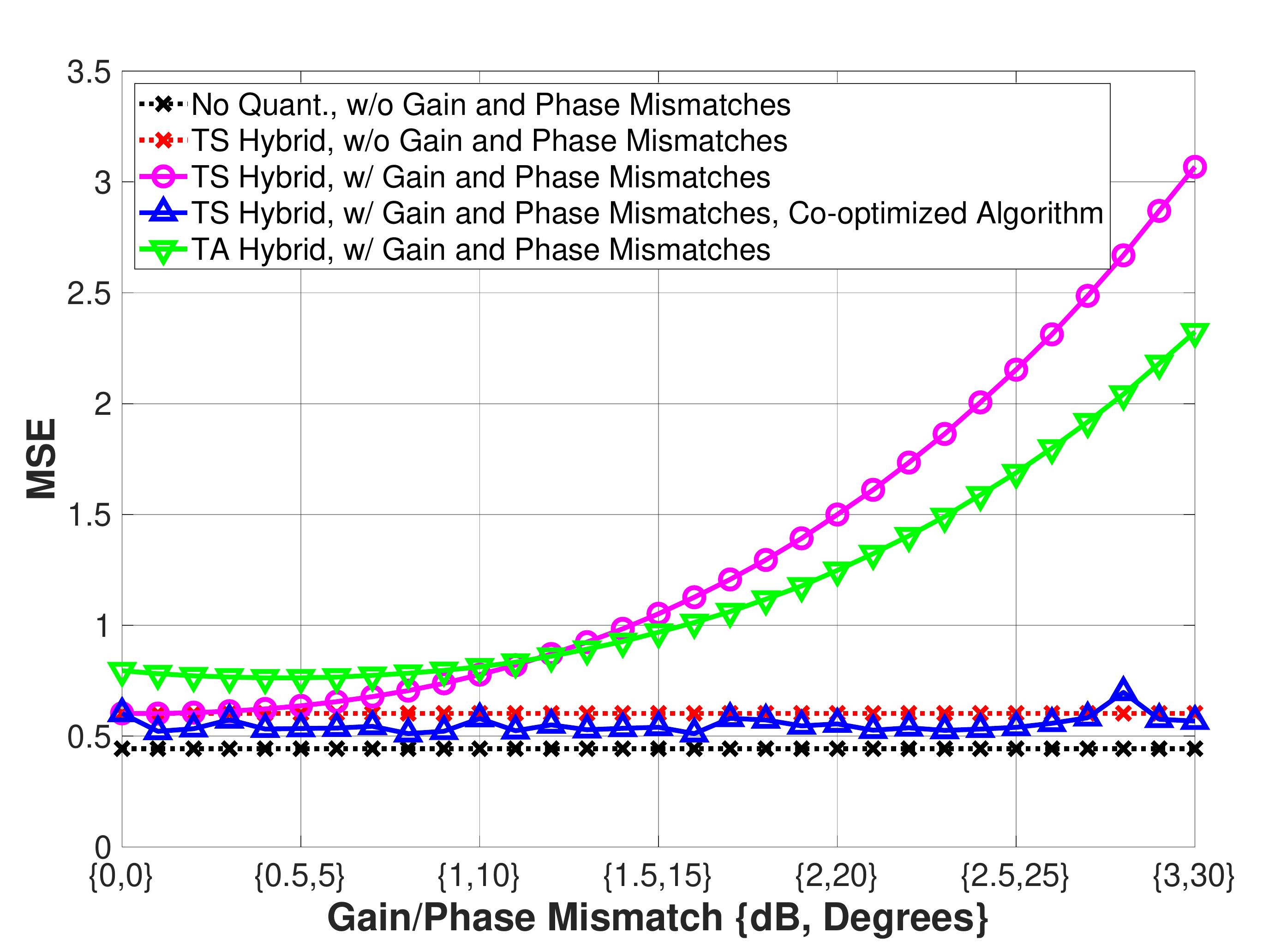}
    \caption{MSE  vs. gain and phase errors $(\alpha, \beta) = (-\zeta, -\delta) = (-2\eta, 2\rho)$ (SNR $= 0$ dB, overall number of bits = 16, 4-bit VM, 25$\%$ sparse).}
    \label{fig:TB_MIMO_sweepImbalances}
\end{figure}

In numerical evaluations, we denote the mean of mismatch parameters as $(\alpha, \beta) \triangleq (\E\{\alpha_{p,n}\}, \E\{\beta_{p,n}\})$, $(\zeta, \delta) \triangleq (\E\{\zeta_{p,n}\}, \E\{\delta_{p,n}\})$ and $ (\eta, \rho) \triangleq (\E\{\eta_{p,n}\}, \E\{\rho_{p,n}\})$ for every $n = \{1, 2, ..., N\}, p = \{1, 2, ..., P\}$ under the hardware-compliant model discussed in \ref{ssec:HardwareMismatches}. The random mismatches are thus i.i.d. among all \acp{vm}. The MSE performance versus the mean gain and phase mismatches $(\alpha, \beta)$ is reported in Fig.~\ref{fig:TB_MIMO_sweepImbalances}, where we set mismatch pairs to be $(\alpha, \beta) = (-\zeta, -\delta) = (-2\eta, 2\rho)$. 
We observe in Fig.~\ref{fig:TB_MIMO_sweepImbalances} that the proposed method clearly eliminates the MSE degradation due to hardware non-idealities. It is seen that for every mismatch level, the co-optimized algorithm achieves the performance of a non-mismatched hybrid \ac{mimo} system where mismatch unaware designs perform substantially worse in recovery.

\subsection{Task-Specific Beamforming and Array Factors}
\label{ssec:Beampatterns}
Algorithm~\ref{alg:Algo1} designs the analog combiner to reject the spatial interferers in analog, in a joint manner with the main system task of recovering $\myVec{s}$ in digital. To evaluate and illustrate the desired suppression of interferers in analog, we next evaluate the receiver \ac{af}. The \ac{af} is a measure of a MIMO receiver gain as a function of an incoming signal's angular direction. An $N\times P$ \ac{mimo} receiver can generate $P$ independent beams directed towards specific angles. 
The \ac{af} generated by $\myMat{A}$ at the $p^{th}$ RF chain is computed as~\cite{soer_JSSC11_switchcap}
\begin{equation}
{\rm AF}_p(\psi) = \sum_{n=1}^{N}[{\myMat{A}}]_{p,n}e^{j\pi n\sin(\psi)}.
\end{equation}

\begin{figure}
\centering
  \subfloat[]{
       \includegraphics[width=0.45\linewidth]{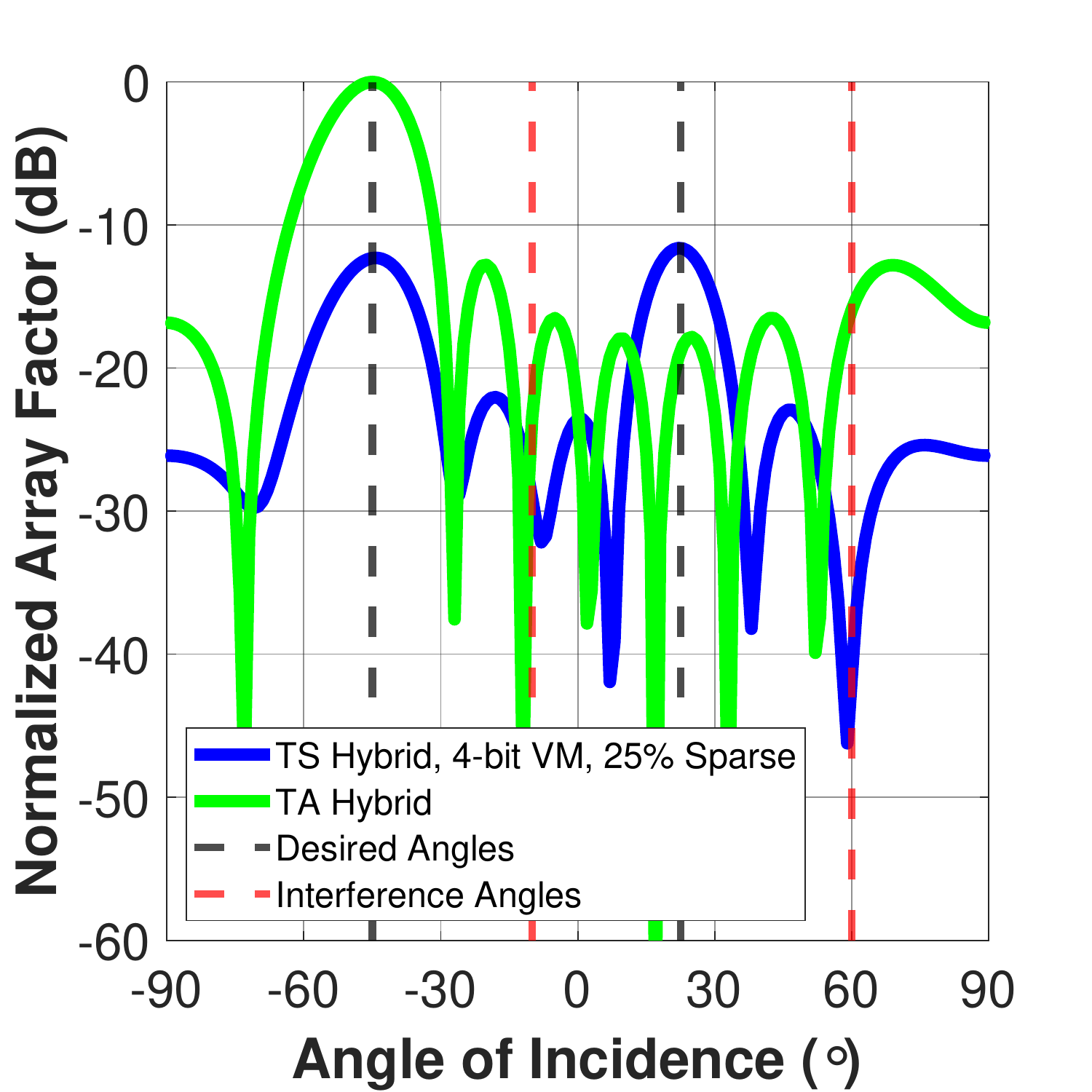}
       \label{fig:setup1_af_1}}
    \hfill
  \subfloat[]{
        \includegraphics[width=0.45\linewidth]{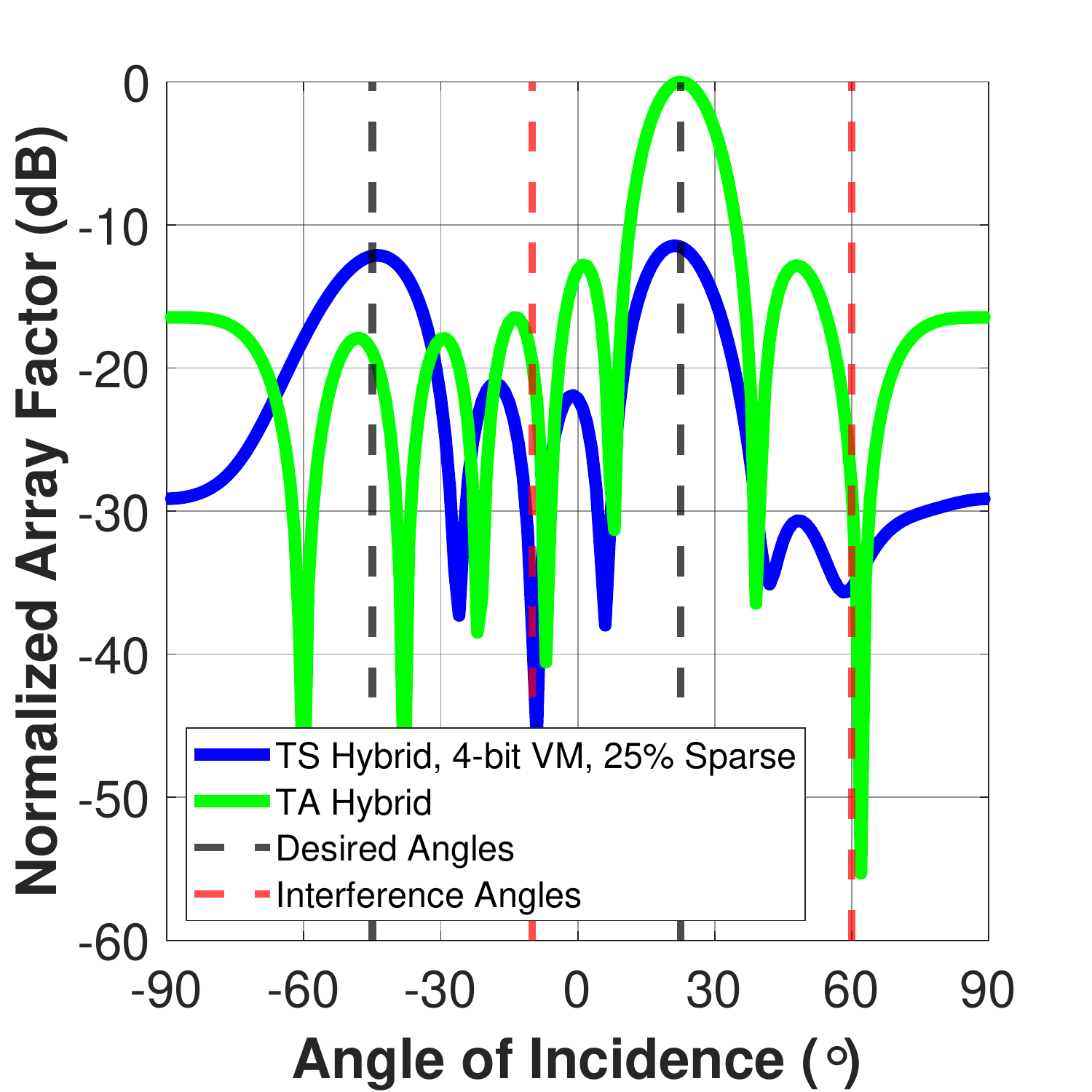} \label{fig:setup1_af_2}}
  \caption{AF for $P=2$ where $K=2$ desired signals at angles $\theta_1= \pi/8$ $\theta_2= -\pi/4$ and $M=2$  interferers at angles $\phi_1= -\pi/18$ $\phi_2= \pi/3$ (SNR = 0 dB, overall 16 bits): (a) first output; (b) second output.} 
  \label{fig:setup1_af_tot} 
\end{figure}

\begin{figure}
\centering
  \subfloat[]{
       \includegraphics[width=0.45\linewidth]{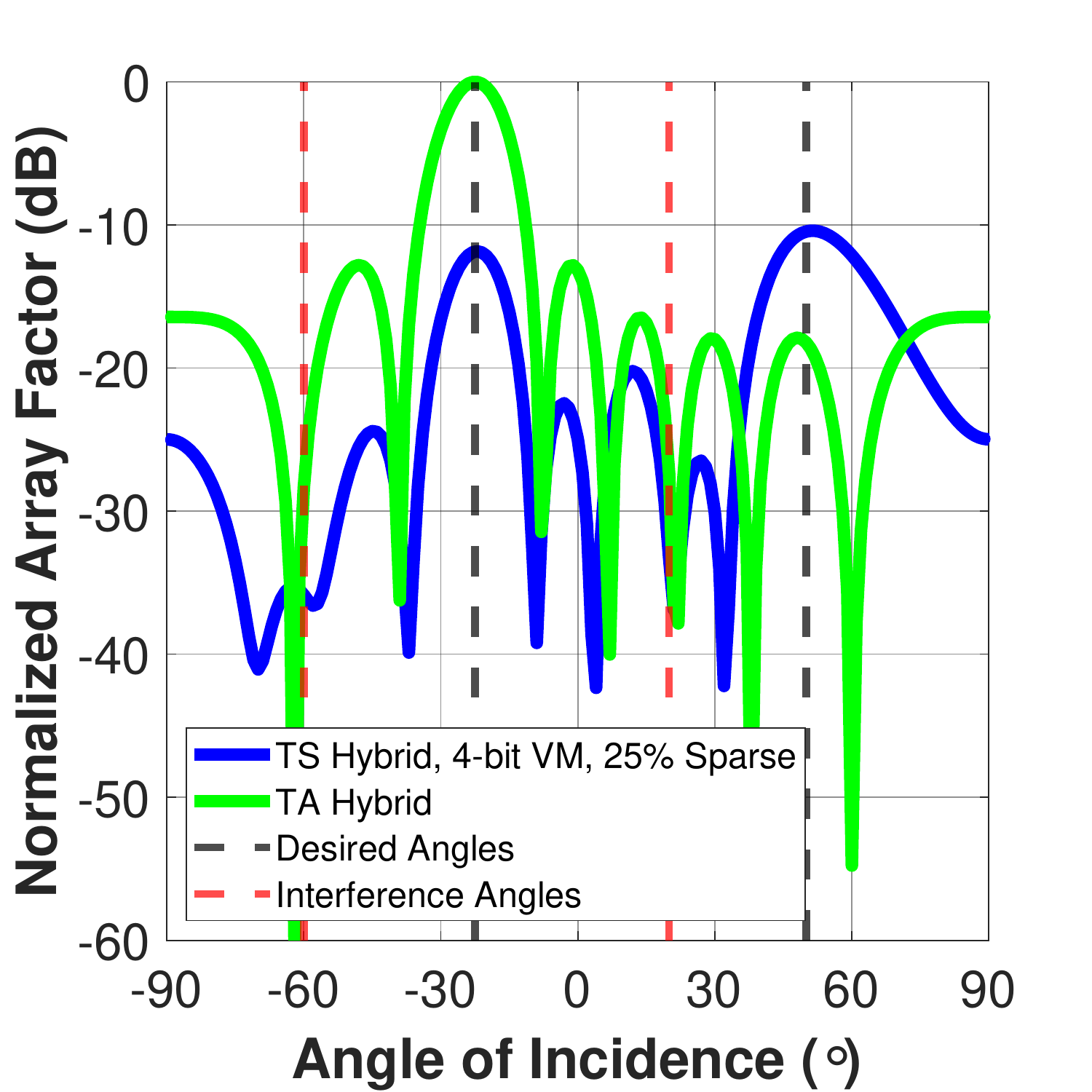}
       \label{fig:setup2_af_1}}
    \hfill
  \subfloat[]{
        \includegraphics[width=0.45\linewidth]{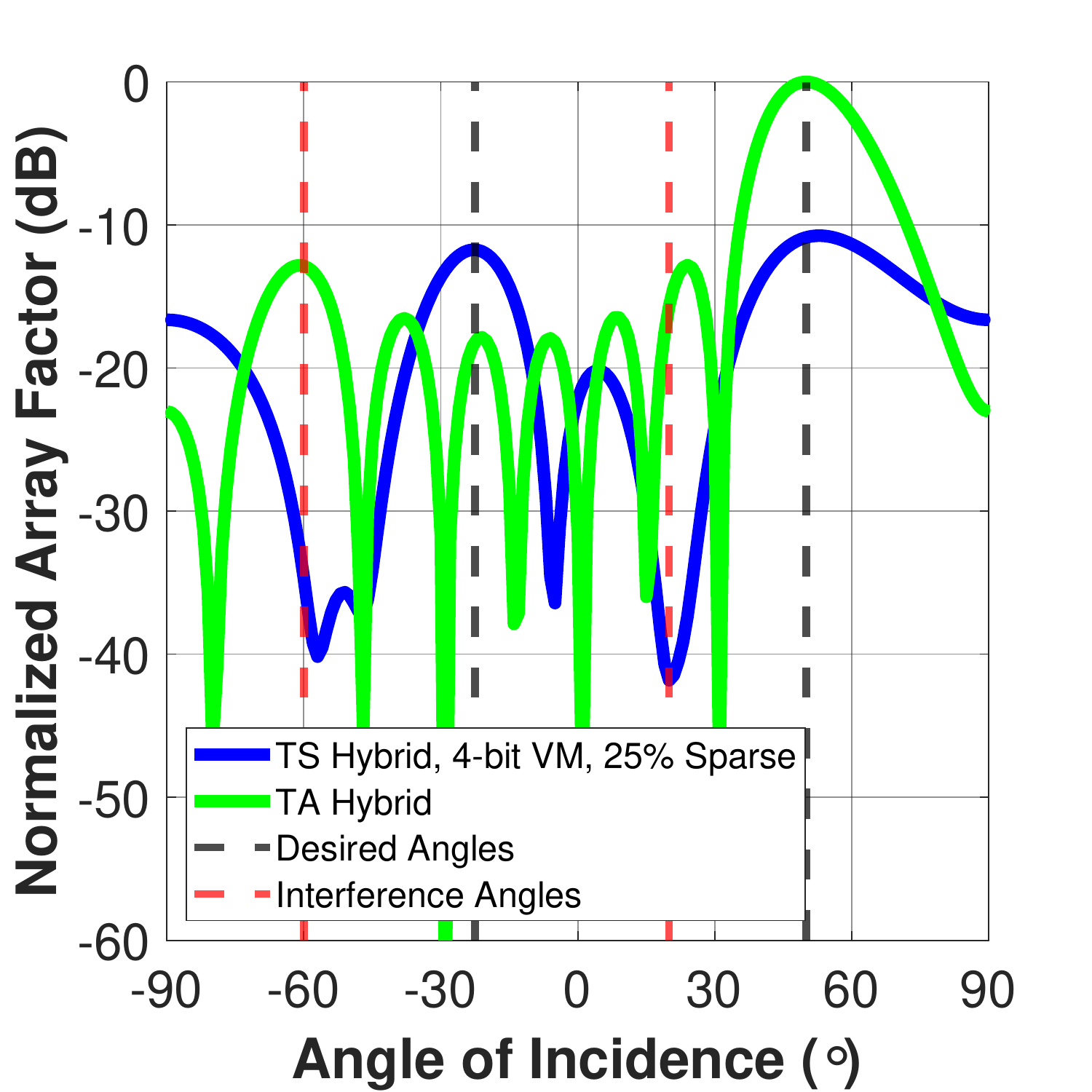}
        \label{fig:setup2_af_2}}
  \caption{AF for $P=2$ where $K=2$ desired signals at angles $\theta_1= -\pi/8$ $\theta_2 = 5\pi/18$ and $M = 2$  interferers at angles $\phi_1 = -\pi/3$ $\phi_2 = \pi/9$ (SNR = 0 dB, overall  16 bits): (a) first output; (b) second output.} 
  \label{fig:setup2_af_tot} 
\end{figure}

In addition to the setup described in Subsection~\ref{ssec:SimSetup}, we also consider another configuration with different angle realizations. The second setup consists of $K=2$ desired signals at angles $\theta_1=-\frac{\pi}{8}, \theta_2=\frac{5\pi}{18}$, and $M=2$ interferers at angles $\phi_1=-\frac{\pi}{3}, \phi_2=\frac{\pi}{9}$, all having same variances as the first setup. The AF computed at the $P=2$ analog outputs of the proposed system is illustrated in Fig.~\ref{fig:setup1_af_tot} and Fig.~\ref{fig:setup2_af_tot} where SNR level and the overall number of bits are set to 0 dB and 16 bits, respectively. It is compared with a task-agnostic conventional hybrid \ac{mimo} receiver whose analog combiner beamforms towards $\theta_1$ (Fig.~\ref{fig:setup1_af_1}-\ref{fig:setup2_af_1}) and $\theta_2$ (Fig.~\ref{fig:setup1_af_2}-\ref{fig:setup2_af_2}). The beam patterns achieved by Algorithm~\ref{alg:Algo1} are directed towards both the desired angles $\theta_{1}$ and $\theta_{2}$, forming a linear combination of the desired signals at the output, while suppressing the interferers at angles $\phi_{1}$, $\phi_2$ by $\ge36$dB. However, as observed in Fig.~\ref{fig:setup1_af_tot}, since the analog combiner is not only optimized for beamforming but rather designed to facilitate recovery from quantized observations, our analog combiner achieves lower AF gain for the desired signals compared to the conventional beamforming. Nonetheless, the lower AF gain does not harm the task-specific recovery accuracy as shown in Fig.~\ref{fig:TB_MIMO_sweepBitsUnmismatched}.

\begin{figure}
\centering
  \subfloat[]{
       \includegraphics[width=0.45\linewidth]{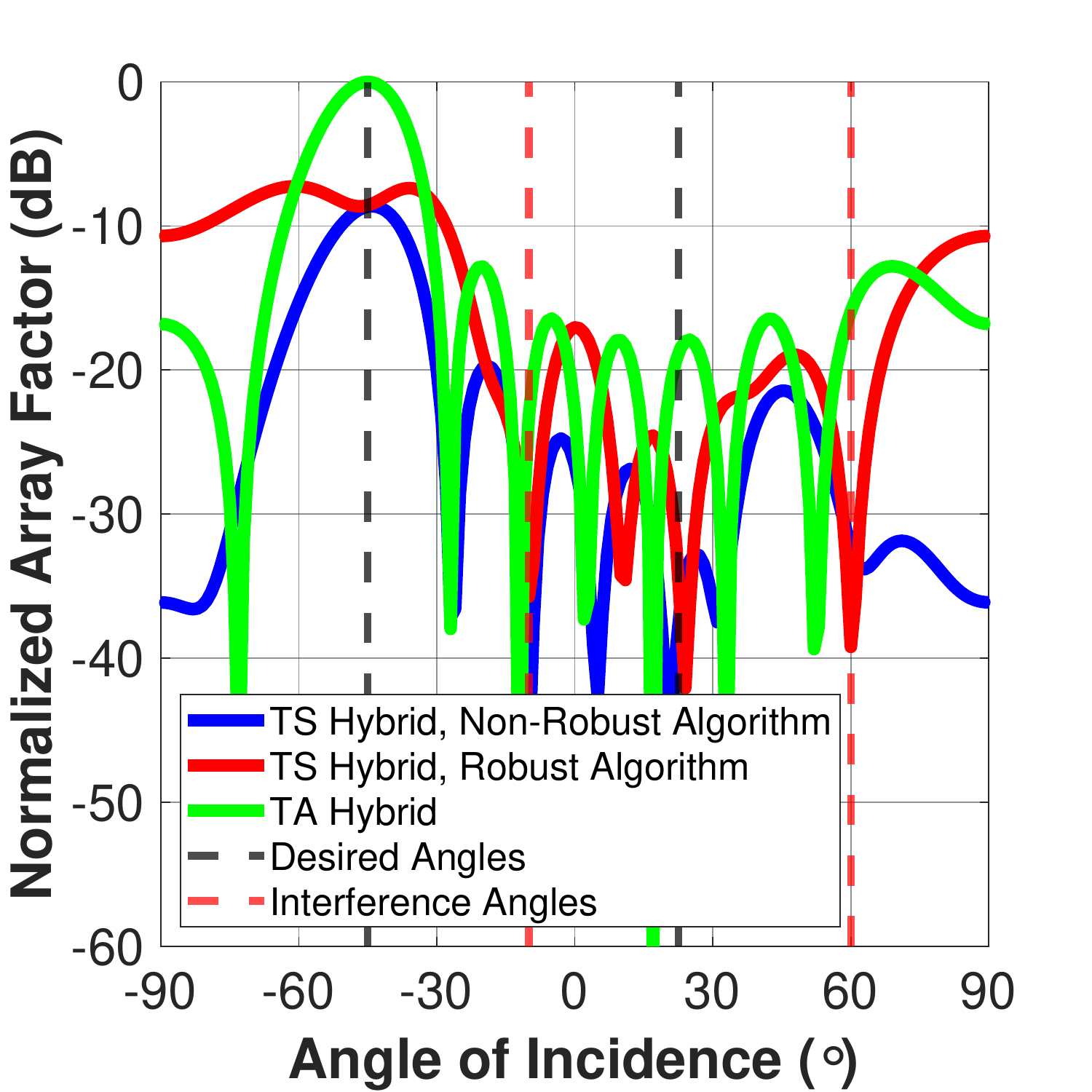}
       \label{fig:AoA_af_1}}
    \hfill
  \subfloat[]{
        \includegraphics[width=0.45\linewidth]{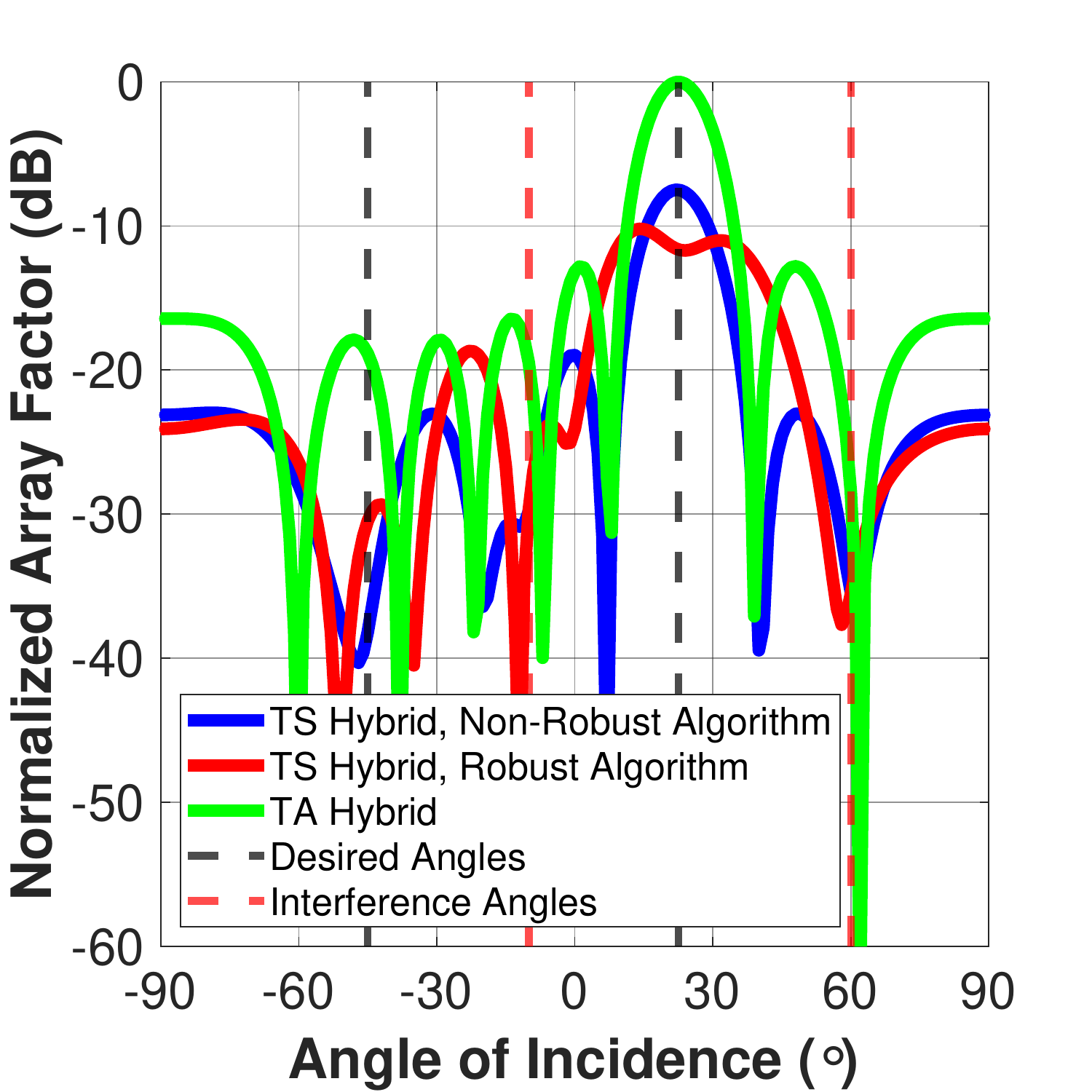}
        \label{fig:AoA_af_2}}
  \caption{AF for $P=2$ where $K=2$ desired signals at angles $\theta_1= \pi/8$ $\theta_2= -\pi/4$ and $M=2$  interferers at angles $\phi_1= -\pi/18$ $\phi_2= \pi/3$ (SNR = 0 dB, overall 16 bits, error margin $\epsilon = 5^\circ$, 4-bit VM, 0$\%$ sparse): (a) first output; (b) second output.} 
  \label{fig:AoA_af_tot} 
\end{figure}

Next, we show the AF computed at the $P=2$ analog outputs of the robust Algorithm~\ref{alg:Algo2} operating with inaccurate \acp{aoa} with error margin $\epsilon = 5^{\circ}$ in Fig.~\ref{fig:AoA_af_tot}, and compare it with non-robust task-specific hybrid \ac{mimo} as well as task-agnostic conventional hybrid \ac{mimo} receiver. The robust analog combiner beamforms towards angles $\theta_{1}$(Fig.~\ref{fig:AoA_af_1}) and $\theta_{2}$(Fig.~\ref{fig:AoA_af_2}). It is observed that the peaks of the main lobes are flattened which enables better accuracy within the range possible \acp{aoa} while still being able to suppress the interferers at angles $\phi_{1}$, $\phi_2$ by $\ge30$dB. 

\subsection{Power Consumption Model}
\label{ssec:PowerConsumption}
We conclude our numerical study by estimating the power consumption of the proposed hybrid \ac{mimo} receiver and the benchmark receivers. To that aim, we use the measured power consumption of  hardware components reported in existing state-of-the-art integrated designs~\cite{soer_JSSC17_4beam, ellinger2009vm, ragaie09mixer, rebeiz17mixer, mendez2016hybrid, lee10badc, sodini2008ADC}. 

Accordingly, the power consumption of an $N\times N$ fully-digital \ac{mimo} receiver is estimated by: 
\begin{equation*}
    P_{\rm FD} = N\cdot P_{\rm LNA} + N \cdot P_{\rm MIX} + 2 \cdot N \cdot P_{\rm BB} + 2 \cdot N\cdot P_{\rm ADC}.
\end{equation*}
Here, $P_{\rm LNA}$ is the power consumed by a low-noise amplifier, $P_{\rm MIX}$ is the power of the mixer, and $P_{\rm BB}$ and $P_{\rm ADC}$ are baseband amplifier and ADC power consumption, respectively, doubled for I and Q paths.

The power consumed by an $N\times P$ hybrid \ac{mimo} receiver is estimated by:
\begin{equation*}
    P_{\rm HYB} = \gamma_{SP} \cdot N \cdot P \cdot P_{\rm VM} + P \cdot P_{\rm MIX} + 2 \cdot P\cdot P_{\rm BB} + 2 \cdot P\cdot P_{\rm ADC}.
\end{equation*}
Here, $P_{\rm VM}$ is the power consumed by a low-noise \ac{vm}-amplifier and $\gamma_{SP}$ is the analog combiner sparsity coefficient: $\gamma_{SP} = 1$ denotes a non-sparse $\myMat{A}$, while $\gamma_{SP} = 0.75$ corresponds to 25$\%$ sparsity. 

The estimated power consumption of each hardware component and the total power consumption of task-specific and -agnostic MIMO receivers are summarized in Table~\ref{table:power_cons}. The power scaling for various quantization levels of the VMs is based on \cite{xum2019vmFoM} when using 8 bits for high-resolution VMs. For the ADC power estimation, we utilize Walden FoM~\cite{cornell2020power, sodini2008ADC}. Our results show that the proposed power-saving techniques, e.g., 25$\%$ sparsity, 4-bit VMs, 4-bit ADCs, provide more than 58$\%$ reduction in power compared to the task-agnostic \ac{mimo} architectures using high-resolution ADCs, high-power LNAs, and high-resolution VMs. These notable power gains add to the improved \ac{mse} performance shown and the spatial interferer rejection observed in the previous subsections.
\begin{table}
{\footnotesize 
\begin{center}
  \caption{Estimated power consumption comparison.}
  \label{table:power_cons}
\begin{tabular}{ |c|c| } 
 \hline
Hardware Component / System & Power (mW)\\ 
 \hline
  LNA/VM $ P_{\rm LNA/VM}$ 
  (1-5 GHz 8 bit/4 bit) \cite{soer_JSSC17_4beam, ellinger2009vm} & 20/10 \\ 
 Mixer with LO Gen (1-5 GHz) $P_{\rm MIX}$  \cite{ragaie09mixer, rebeiz17mixer} & 15 \\ 
 Baseband Amplifier $P_{\rm BB}$ \cite{mendez2016hybrid} & 5 \\ 
 ADC $P_{\rm ADC}$ (100 MS/s 10 bit/4 bit) \cite{lee10badc, chandrakasan06} & 10/0.5 \\ 
 \hline
 Fully-Digital \ac{mimo} Receiver ($8\times8$) & 520 \\ 
 Conventional Hybrid \ac{mimo} Receiver ($8\times2$) & 410 \\ 
 \textbf{Task-Specific Hybrid \ac{mimo} Receiver ($\mathbf{8\times2}$)} & \textbf{172} \\ 
  \hline
\end{tabular}
\end{center}
}
\end{table}

\section{Conclusion}
\label{sec:Conclusions}
In this work, we studied a power-efficient hybrid \ac{mimo} receiver design with embedded beamforming and low-resolution ADCs using task-specific quantization under certain practical non-idealities. We introduced a power-efficient analog and digital joint optimization framework, incorporating sparse analog combining and considering the finite resolution of the configurable analog hardware. Furthermore, we proposed a robust joint optimization algorithm to cope with mismatched \acp{aoa}, and extended our design to handle phase and gain errors. 
The proposed hybrid \ac{mimo} receiver notably outperforms the task-agnostic \ac{mimo} receivers by achieving improved MSE performance and successfully suppressing undesired spatial interferers at lower power and lower quantization rate, while being robust to common mismatches.

\appendix
\subsection{Proof of Lemma~\ref{lem:mseAoA}}
\label{proof:lem2}
The \ac{mse} expression in \eqref{eqn:MSEdecomposed} can be formulated as
\begin{align}
    {\rm MSE} &= \E\{\| \myVec{s} - \myMat{B}\mathcal{Q} (\myMat{A}\myVec{x})\|^2\} 
    \stackrel{(a)}{=} \E\|\myVec{s} - \myMat{B}(\myMat{A}\myVec{x} + \myVec{e})\|^2\} \notag \\
    &\stackrel{(b)}{=} \E \{\|\myVec{s} - \myMat{B}(\myMat{A}\myVec{x})\|^2\} + \E \{\|\myMat{B}\myVec{e}\|^2\}. 
\end{align}
Here, $(a)$ follows from \cite[Thm. 2]{gray93Dither} when ${\rm Pr}(|(\myMat{A}\myVec{x}_l + d_l)| > \gamma) = 0$, where $\gamma$ is the dynamic range of \acp{adc} and $d_l$ is the dither signal. Namely, the sum of the input to the \acp{adc} and dither signal is always within the dynamic range, thus the output of the \acp{adc} can be written as $\myMat{A} \myVec{x} + \myVec{e}$ where $\myVec{e}$ is the quantization noise. Furthermore, by \cite[Thm. 1]{shlezinger2019hardware}, $\myVec{e}$ is uncorrelated with observation $\myVec{x}$, and the desired signal $\myVec{s}$, i.e., $\E \{\myVec{xe}^H\} = 0$ and $\E \{\myVec{se}^H\} = 0$, which implies $(b)$, and has uncorrelated zero-mean entries with variance $\frac{\Delta_p}{6}^2$, where $\Delta_p = \frac{2\gamma}{b}$ is the quantization spacing. 

By distributing each term and using the linearity and cyclic property of the trace operator we have
    ${\rm MSE} = {\rm Tr} \Big( \CovMat{s} - 2\CovMat{sx}\myMat{A}^H\myMat{B}^H 
    + \myMat{BA}\CovMat{x}\myMat{A}^H\myMat{B}^H + 
    \myMat{B}\frac{\Delta_p^2}{6}\myMat{B}^H \Big)$.
Finally, setting $\kappa := \eta^2(1 -\frac{\eta^2}{3b^2})^{-1}$ where $\eta$ is a coefficient tuned to guarantee negligible overloading probability of the ADCs, and adapting \cite[Thm. 1]{shlezinger2019hardware}, yields
\begin{align*}
    {\rm MSE} &= {\rm Tr}\Bigg( {\CovMat{\myVec{s}}} -2 \CovMat{\myVec{s}\myVec{x}}\myMat{A}^H\myMat{B}^H \notag \\
    &+ \myMat{B}\myMat{A}\CovMat{\myVec{x}}\myMat{A}^H\myMat{B}^H + \myMat{B}\frac{2\kappa \cdot {\rm Tr}(\myMat{A} \CovMat{\myVec{x}} \myMat{A}^H)}{3b^2 \cdot P}\myMat{B}^H\Bigg),
\end{align*}
thus proving the lemma.

\bibliographystyle{IEEEtran}
\bibliography{IEEEabrv,refs}

\end{document}